\documentclass[acmsmall]{acmart}
\AtBeginDocument{%
  }

\setcopyright{rightsretained}
\acmJournal{PACMHCI}
\acmYear{2025} \acmVolume{9} \acmNumber{2} \acmArticle{CSCW181}
\acmMonth{4} \acmDOI{10.1145/3711079}

\usepackage{enumitem}
\usepackage{multirow} 
\usepackage{url} 
\usepackage{graphicx} 

\begin{document}

\title[Systematic Literature Review]{A Systematic Literature Review on Equity and Technology in HCI and Fairness: Navigating the Complexities and Nuances of Equity Research}
\author{Seyun Kim} 
\orcid{0000-0001-5905-7093}
\email{seyunkim@cs.cmu.edu}
\affiliation{%
  \institution{Carnegie Mellon University}
  \city{Pittsburgh}
  \state{Pennsylvania}
  \country{USA}
}

\author{Yuanchen Bai} 
\orcid{0009-0004-2140-7894}
\email{yb299@cornell.edu}
\affiliation{%
  \institution{Cornell University}
  \city{Ithaca}
  \state{New York}
  \country{USA}
}

\author{Haiyi Zhu \textsuperscript{*}}
\orcid{0000-0001-7271-9100}
\email{haiyiz@cs.cmu.edu}
\affiliation{%
  \institution{Carnegie Mellon University}
  \city{Pittsburgh}
  \state{Pennsylvania}
  \country{USA}
} 

\author{Motahhare Eslami  \textsuperscript{*}}
\orcid{0000-0002-1499-3045}
\email{meslami@andrew.cmu.edu}
\affiliation{%
  \institution{Carnegie Mellon University}
  \city{Pittsburgh}
  \state{Pennsylvania}
  \country{USA}
} 

\renewcommand{\thefootnote}{\fnsymbol{footnote}}
\footnotetext[1]{These authors contributed equally}

\renewcommand{\shortauthors}{Seyun Kim et al.}

\begin{abstract}
Equity is crucial to the ethical implications in technology development. However, implementing equity in practice comes with complexities and nuances. In response, the research community, especially the human-computer interaction (HCI) and Fairness community, has endeavored to integrate equity into technology design, addressing issues of societal inequities. With such increasing efforts, it is yet unclear why and how researchers discuss equity and its integration into technology, what research has been conducted, and what gaps need to be addressed. We conducted a systematic literature review on equity and technology, 
collecting and analyzing 202 papers published in HCI and Fairness-focused venues. Amidst the substantial growth of relevant publications within the past four years, we deliver three main contributions: (1) we elaborate a comprehensive understanding researchers' motivations for studying equity and technology, (2) we illustrate the different equity definitions and frameworks utilized to discuss equity, (3) we characterize the key themes addressing interventions as well as tensions and trade-offs when advancing and integrating equity to technology. Based on our findings, we elaborate an equity framework for researchers who seek to address existing gaps and advance equity in technology. 
\end{abstract}


\received{January 2024}
\received[revised]{July 2024}
\received[accepted]{October 2024}

\maketitle

\section{Introduction}
As concerns over ethical implications in the development of technologies grow, \textit{equity} has emerged as a key factor in discussing and tackling these challenges. One way of describing equity is acknowledging the existence of an uneven playing field due to historical patterns of injustice and disparities in resource allocations~\cite{jones2009equity, hope2010inquiry, murphy1984factors}. 
Promoting equity in practice, however, is quite challenging. For instance, race-based affirmative action in college admissions~\cite{supremecourt}, which was originally designed to provide opportunities to underrepresented groups, has intensified the debate over equity's definition and implementation. While some view the Supreme Court's ruling against affirmative action in college admissions as a step away from demographic biases in the admission processes, others fear it exacerbates existing inequities, particularly affecting historically disadvantaged applicants. The implications of this ruling extend into the realm of technology, notably raising questions about the legality and ethical considerations of algorithmic affirmative action, which similarly seeks to foster equity in automated systems~\cite{algorithmic_affirmativeAction}.

To tackle such challenges, recent years have witnessed growing efforts in studying and incorporating equity into the technology design and development in the research community. This is particularly evident in fields such as Human-Computer Interaction (HCI) and fairness, when researchers study and examine the integration of equity into technology and computing systems in different domains such as education, public sector, and non-profits~\cite{lee2019webuildai, robertson2021modeling, harrington2019deconstructing, black2022algorithmic}. We have also seen the emergence of new venues of relevant topics such as conferences focused on fairness \& ethics, including the ACM Conference on Fairness, Accountability, and Transparency (FAccT), the AAAI/ACM Conference on AI, Ethics and Society (AIES), and the ACM conference on Equity and Access in Algorithms, Mechanisms, and Optimization (EAAMO), as well as workshops and panels on fairness, social justice and equity such as the workshop on Ethical Pluralism in CSCW~\cite{rifat2023many}, the Fair and Responsible AI workshop at CHI~\cite{lee2020human}, and the Equity \& Inclusivity workshop at IDC~\cite{fox2016exploring, sobel2017equity}. However, the role of equity in these emerging fields remains unclear. It remains unclear as to why and how researchers advance equity and its integration into technology, as well as what research has been conducted and what gaps need to be addressed. 

To address these questions, we conducted a systematic literature review on equity research and its integration into technology, using a broad and open-ended definition of technology. Literature reviews are effective for providing insights and identifying trends, key themes and gaps in the research community~\cite{kim2021human,frohlich2022blockchain,gray2023mapping, mehrabi2021survey, wieringa2020account}. In this paper, we examined 202 papers across various research venues from their inception to 2023, representing two interconnected fields of study: HCI (e.g., CHI, CSCW, DIS, GROUP) and Fairness-focused venues (e.g., FAccT, AIES, EAAMO). 

Our investigation revealed a scarcity of equity-related research prior to 2018 in both HCI and Fairness-focused venues, characterized by fewer than five papers published annually on this topic. However, we observed a noticeable surge in the past four years, with research articles growing up to sixfold annually. This increased interest in equity and its integration into technology aligns with the emergence of Fairness-focused venues in recent years. Our analysis of this growing dataset offers the following contributions:  

\begin{itemize}

\item \textit{WHAT Shapes Equity Research?}
We found a significant discrepancy in the clarity and uniformity of how equity is defined in academic research, with only 27\% of our dataset explicitly clarifying the term. For papers that do define equity, we pinpointed three categories of definitions (need-based, contribution-based, and equality-based), and discussed the differences between them. The lack of an explicit and clear equity definition, especially evident in HCI studies compared to those centered on Fairness, may hinder the development of a common framework and shared understanding of the equity concept within the academic community. 
We also identified the primary frameworks researchers draw upon to discuss equity, including social-justice frameworks, Feminist HCI, and ableism. 

\item \textit{WHY Study Equity and Technology?} Our analysis indicates that the existence of ingrained inequities in both technology and wider society serves as a primary catalyst for researchers to delve into equity studies. This motivation stems from a need to address and understand various forms of discrimination and challenges, such as the digital divide and health disparities, particularly affecting marginalized communities, including immigrants, LGBTQ+ groups, and people of color. 
Another significant impetus for this research area arises from the inadequacy and shortcomings of prior efforts in effectively addressing equity objectives. This gap has motivated researchers to explore more effective and inclusive approaches, seeking to develop methodologies and frameworks that better align with the diverse needs and experiences of underrepresented groups.

\item \textit{HOW Do Researchers Conduct Equity Research?}
Our analysis revealed several approaches researchers take in studying the interaction of equity and technology. We found that a subset of the papers in our dataset treats equity primarily as a motivational backdrop for their research without deeply engaging in its study or advancement.
In contrast, the rest of the papers with a focus on equity are dedicated to either identifying and surfacing existing inequities in current technologies or devising intervention strategies to infuse equity into technology design and development. These interventions include a) creating systems or algorithms aimed at enhancing equitable opportunities for underrepresented and marginalized groups, b) implementing interventions based on frameworks, and c) incorporating participatory processes with stakeholders to achieve more equitable technology designs.

\item \textit{Tensions and Trade-offs in Advancing Equity}.
In identifying the ways researchers conduct equity research, we also identified various tensions and trade-offs between equity and other competing values, including utility-focused values (e.g., efficiency and accuracy), social values (e.g., equality and privacy), and sometimes, conflicting aspects of equity itself.

\item \textit{Developing an Equity Framework.} Based on the findings illustrated above, we developed an equity framework with four dimensions for researchers to reflect on when integrating and advancing equity. These four dimensions include understanding what, how, and why researchers conduct equity research, followed by discussions on tensions and trade-offs of values. We provide several guidelines and recommendations regarding each of the equity framework dimensions that researchers can consider when studying the intersection of equity and technology. These guidelines include providing a clear and explicit equity definition, 
identifying the related lenses and frameworks when studying equity, 
going beyond bias mitigation, 
bridging different efforts in fostering equity, 
enhancing value transparency in advancing equity, and
reflecting on technology as a solution. We discuss each of these recommendations in the Discussion section.

\end{itemize} 

We conclude by discussing the need for researchers to reflect on whether technology is an adequate solution for equity challenges. We encourage researchers to critically evaluate how unintentional inequities emerged from advancing equity with and through technologies. 

\section{Related Work} 

\label{relatedwork} 
To help contextualize this literature review, we further elaborate on different conceptualizations of equity as well as equity-related concepts such as social justice, fairness, feminist HCI, and methods involving stakeholders. We illustrate a review of equity-related literature reviews and how our work builds on this line of work. 

\subsection{Background on Equity}
Equity has been defined and conceptualized differently across various fields, reflecting its multifaceted nature. An early definition of equity traces back to the context of legal justice, where Aristotle describes equitable decisions as those in which a judge transcends the law to achieve fairness~\cite{beever2004aristotle}. In this conceptualization, equity is viewed as \textit{distributive justice}, where rewards are allocated based on individual merit~\cite{walster1975equity}.

In addition to Aristotle's definition of equity, \textit{Equity theory}, a concept in the workplace and behavioral psychology, assesses the balance between the contributions of an individual and the rewards they receive \cite{pritchard1969equity, walster1973new, adams1963towards}. J. Stacey Adams, a behavioral psychologist, developed \textit{Equity Theory} to ensure fair treatment for employees by evaluating individuals' merit using factors such as time and ability~\cite{adams1963towards}. When individuals find that their pay or reward do not align with their merit, they perceive their circumstances as less fair, impacting their motivations, job satisfaction, and performance~\cite{al2012utility}.

While equity theory focuses on ``merit'', other conceptualizations of equity focus on ``needs'' as the basis for distributing resources. For example, a well-known example illustrates three kids: a tall kid, a short kid, and a kid in a wheelchair, trying to watch a baseball game over a fence\footnote{\url{https://www.anchor.org.uk/media/press-room/equity-vs-equality-why-should-we-focus-equity}}. The tallest kid can easily see the game over the fence, while the shortest kid and the kid in a wheelchair cannot. In this scenario, equity recognizes that everyone receives the resources they need to be successful. The tallest kid does not need any resources to see the game, the shortest child stands on two boxes, and the kid with the wheelchair receives a ramp instead of boxes to see the game over the fence.

While previous conceptualizations of equity consider individual differences as a basis for distributing resources, some equate equity with equality, where everyone receives the same resources (e.g., all the kids in the example above receive two boxes). For example, in an investigation of inequities within social systems, the authors stated that ``equity is achieved when a societal system produces \textit{equal statistics} across groups''~\cite{reader2022models}. These variances and discrepancies in the concept of equity have led us to investigate how equity is defined and utilized in interaction with technology. 

\subsection{Concepts Related to Equity}
We cannot discuss equity without discussing relevant themes such as social justice, feminism, and fairness. As HCI grew as a field to address and surface complex societal challenges, integrating and advancing these concepts in the design of technology became important for developing ethical, responsible, and just technologies. Hence, understanding the discourse around these themes could help the research community develop a comprehensive picture on equity.

\subsubsection{Social Justice.} Similar to equity, there is no universal definition or method for pursuing social justice~\cite{chordia2024social}. According to John Rawls, justice ensures that everyone has equal rights, with the greatest benefit and resources going to those who are the least advantaged~\cite{mandle2009rawls}. Additionally, L\"otter describes social justice as a multifaceted concept, elaborating six dimensions of social justice: recognition, reciprocity, enablement, distribution, accountability, and transformation~\cite{dombrowski2016social, fraser2008social}. These dimensions have inspired HCI researchers to develop social justice-oriented interaction design, helping designers to address societal challenges (e.g., marginalized experiences, systematic discrimination) and develop ethical design~\cite{fox2016exploring, chordia2024social}. HCI researchers have extended their pursuit for justice beyond the outcome of the technology and research by committing to justice in the methodology and the process of research~\cite{chordia2024social}. These examples include committing to integrating the voices of impacted stakeholders and distributing the power to marginalized communities in the design and research process~\cite{harrington2019deconstructing}. 

\subsubsection{Feminist HCI} In parallel, HCI scholars have integrated feminism into interaction design research and practices, creating what is known as \textit{feminist HCI}~\cite{bardzell2010feminist, bardzell2011towards, fiesler2016archive}. Feminism as a framework critically analyzes ``concepts, assumptions, and epistemology of HCI'' as it encourages challenging discussions such as how to prevent perpetuating discrimination against marginalized communities. Feminist HCI proposes six qualities to HCI design: pluralism, participation, advocacy, ecology, embodiment, and self-disclosure~\cite{bardzell2010feminist}. Participation, for example, values the communication between the designers and impacted stakeholders in the process of creating and evaluating design artefacts~\cite{bardzell2010feminist}.

\subsubsection{Participatory Design.} Participatory Design ensures that power and responsibilities are distributed across stakeholders. This distribution assures the voices of underrepresented communities are heard. Since the 1980s, scholars have investigated the work of engaging with impacted stakeholders through Participatory Design (PD). PD originated from Scandinavian union workers who ensured workplace democracy by advocating their rights to participate in developing technologies in their workplaces. PD challenges the existing power structure and surfaces diverse voices into the design processes~\cite{muller1993participatory}. Academic workshops such as “Social Justice and Design: Power and Oppression in Collaborative Systems”~\cite{fox2017social} emerged with the goal for engaging impacted stakeholders and re-distribute the power in research practices.

\subsubsection{Fairness.} The increased attention towards fairness in HCI aligns with the growing expectations of artificial intelligence and machine learning to behave not only accurately but also ethically and responsibly~\cite{mehrabi2021survey}. This growth is evident in the emergence of Fairness-focused venues such as FAccT in 2018, which originated as a workshop at NeurIPS (Neural Information Processing Systems)~\cite{laufer2022four}. Researchers have dedicated to surface how impacted stakeholders perceive fairness in relation to technologies. Prior scholars investigated what factors contribute to either procedural or outcome fairness, acknowledging that fairness is subjective~\cite{ schoeffer2021appropriate, lee2017algorithmic, cheng2022child, cheng2021soliciting, cheng2021soliciting, claure2023social}. For example, Lee et al. explored how Uber drivers perceived the app to be unfair due to the lack of autonomy over the app's structure and decision-making processes~\cite{lee2015working}. Additionally, researchers have approached fairness through computational methods, developing metrics to evaluate the fairness of technologies. These assessments include demographic parity and equalized odds
~\cite{laufer2022four, verma2018fairness, cheng2022child}. While demographic parity ensures that the positive rate is equal to all groups regardless of differences in circumstances between the groups~\cite{raz2021group}, equalized odds guarantee the false positive rate or the false negative rate to be same across all groups~\cite{mutlu2022contrastive}. Due to these differences in fairness metrics, avid discussions on which metric and variable are the most suitable for assessing an algorithm emerged~\cite{abbasi2021fair, jacobs2021measurement, angwin2022machine}. This sparked debates on how ``fairness'' should be measured, as even the same algorithm can be perceived as biased or fair depending on which fairness metric is used for evaluation~\cite{dressel2018accuracy}. 

Despite a number of systematic literature reviews focusing on fairness, social justice, and participatory design~\cite{jacobs2021measurement, mehrabi2021survey, caton2024fairness,bach2024systematic,chordia2024social,duque2019systematic, zahlsen2022challenges}, equity has not been the main focus of these papers. Therefore, as these concepts are related to equity, we investigate how they have been addressed by researchers in this literature review.

\subsection{Reviews of Equity-Related Literature Review}
Systematic literature reviews are the foundation for new knowledge as they identify current trends, gaps, and opportunities for scholars in the research community~\cite{stefanidi2023literature, webster2002analyzing}. Especially in rapidly growing fields such as HCI and fairness, literature reviews help researchers critically examine the overarching field’s direction and potential future research~\cite{kim2021human,frohlich2022blockchain,gray2023mapping, sannon2022privacy}. 
Consequently, we see a growing number of literature reviews on equity-related topics such as fairness, accountability, and transparency in technologies. Examples include investigating ethics in AI~\cite{khan2022ethics}, algorithmic accountability~\cite{wieringa2020account}, and examining research topics in the Fairness community~\cite{laufer2022four}. We found literature reviews on social justice explored in contexts such as education~\cite{mills2016social, wiedeman2002teacher, schulze2017investigating} and health care~\cite{dukhanin2018integrating}. For example, Chordia et al. conducted a systematic literature review on social justice research in HCI~\cite{chordia2024social}. The authors highlighted the current state of justice-oriented research, including the discourse on harms and benefits, tools that researchers leverage to conduct equitable processes, as well as the key considerations to address these harms~\cite{chordia2024social}. Systematic literature reviews on equity have investigated under specific contexts such as public services~\cite{cepiku2021equity, ruijer2023social, anselmi2015equity}, education~\cite{poekert2020leadership, wanti2022determining}, and healthcare~\cite{tao2016impact}. Hence, we see an opportunity to provide a comprehensive analysis of the current landscape in equity and technologies to help guide future researchers on this topic.

\section{Method}
\label{method}
\begin{figure}[h!]
    \centering
    \includegraphics[width=\textwidth]{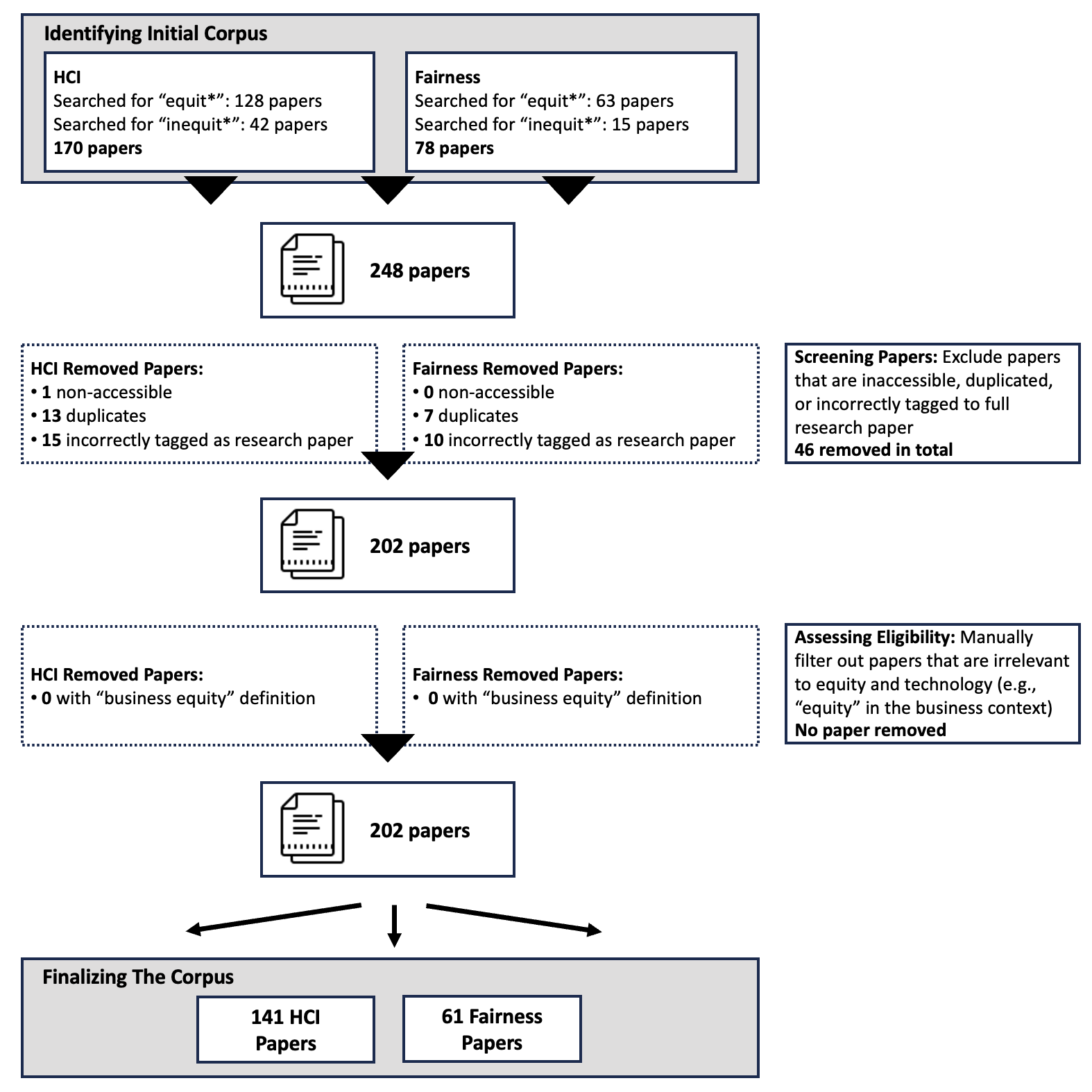}
    \caption{The Preferred Reporting Items for Systematic Reviews and Meta-Analysis (PRISMA) workflow.}
    \Description{The Preferred Reporting Items for Systematic Reviews and Meta-Analysis (PRISMA) workflow, including four main steps: Identifying Initial Corpus, Screening Papers, Assessing Eligibility, and Finalizing The Corpus. We finalized a total of 202 papers in our final corpus.}
    \label{fig:prisma}
\end{figure}

We used the Preferred Reporting Items for Systematic Reviews and Meta-Analysis (PRISMA) workflow~\cite{moher2009preferred} to conduct the systematic literature review. PRISMA consists of four steps: 1) identifying initial corpus 2) screening papers 3) assessing eligibility 4) finalizing the corpus. This method has been used in the research community in domains such as privacy and marginalization~\cite{sannon2022privacy}, algorithmic auditing~\cite{bandy2021problematic}, and algorithmic accountability~\cite{wieringa2020account}. In the following sections, we elaborate the steps involved in creating our corpus which include identifying the initial corpus and screening papers. We elaborate an overview of the PRISMA workflow illustrated in Figure~\ref{fig:prisma}.

\subsection{Identify Initial Corpus} 
To ensure papers on equity and technology are comprehensively represented in our dataset, we created a corpus of papers published in HCI and Fairness-focused venues. We focused on these two research communities, as the HCI community has historically surfaced stakeholders’ needs and voices in the design of technologies and the Fairness community, although nascent, has addressed methods to mitigate bias in technologies. Therefore, we considered these two communities suitable for our systematic literature review. We collected the data in September 2023. The inclusion criteria to identify relevant papers for the final corpus are as follows: 1) the paper is a ``Research Article'', which excludes posters, abstracts, panels, etc. 2) the paper focuses on equity and technology. We did not place any restrictions on the publication date, which will be further elaborated in Section~\ref{overview}. In the following section, we further elaborate on the process of identifying the search terms and the details on the data collection process for the HCI and Fairness-focused venues. 

\textit{HCI.} We used the ACM Digital Library (ACM-DL) and selected the ``sponsored by SIGCHI'' option. SIGCHI (Special Interest Group on Computer-Human Interaction) is a research organization that sponsors or co-sponsors HCI conferences such as CHI, CSCW, and DIS with the goal of advancing HCI. Additionally, we selected the ``SIGACCESS'' option in the ACM-DL. SIGACCESS (Special Interest Group on Accessible Computing) supports researchers committed to ``applying computing and information technologies to empower individuals with disabilities and older adults''~\footnote{https://www.sigaccess.org/}. The International ACM SIGACCESS Conference on Computers and Accessibility (ASSETS) is the flagship conference for SIGACCESS. 

\textit{Fairness-focused venues.} Unlike HCI papers, there is no central database equivalent to ACM-DL for Fairness-focused papers. Therefore, we selected Fairness-focused conferences that do not overlap with ACM SIGCHI and ACM SIGACCESS. The Fairness-focused conferences include the Conference on Fairness, Accountability, and Transparency (ACM FAccT), the Conference on Artificial Intelligence, Ethics, and Society (AIES), the Conference on Equity and Access in Algorithms, Mechanism and Optimization (EAAMO) and the IEEE Conference on Secure and Trustworthy Machine Learning (SaTML). These four conferences were selected based on their interdisciplinary research focusing on the following themes: fairness, accountability, transparency, ethics, equity, and trustworthiness in socio-technical systems and AI.

\subsubsection{Identifying Search Terms} 

We iteratively refined our search terms, capturing papers focused on ``equity'' and ``technology''. This process required testing different keyword combinations and discussions by the research team to finalize the search terms. In our search for equity-relevant papers, we found that they may not explicitly use the term ``equity'' but discuss related topics such as ``accessibility'', fairness', and ``diversity''. We excluded these papers and narrowed down our keyword to ``equity''. This decision was to narrow our focus to papers that explicitly use the word ``equity'', and reduce noise in our corpus.  

Therefore, we initially used ``equity'' and ``technology'' as our search keywords. However, when using these keywords in ACM-DL SIGCHI, the search engine yielded only 5 research articles. Faced with the dearth of papers when using both ``equity'' and ``technology'', we decided to expand our search. We chose ``equit*'' and ``inequit*'' as keywords and removed ``technology''. The choice of (*) was to ensure that we include all forms of ``equity'' and ``inequity''. We removed the term ``technology'' to broaden our research beyond specific technologies. 
\newpage
As a result, the search keywords and parameters we used are as follows: 

{ \fontfamily{qrc} \selectfont 
Title: (equit*) OR 
Abstract: (equit*) OR
Keyword: (equit*) 
} and 

{ \fontfamily{qrc} \selectfont 
Title: (inequit*) OR 
Abstract: (inequit*) OR
Keyword: (inequit*) } 
\\
\\
Search queries with ``equit*'' in titles, abstracts, and/or keywords yielded 128 HCI papers, while queries for ``inequit*'' resulted in 42 HCI papers. We removed 15 papers that were mistakenly tagged as ``Research Article'' by the digital library’s automation tag system, 13 duplicates between the ``equit*'' and ``inequit*'' results, and 1 paper that the authors could not access. After this screening process, the HCI dataset contained 141 papers. The dataset included papers from the following 16 conference venues: ACM Conference on Human Factors in Computing Systems (CHI), ACM Conference on Computer Supported Cooperative Work (CSCW), ACM SIGCHI Conference on Designing Interactive Systems (DIS), ACM International Conference on Supporting Group Work (GROUP), ACM International Conference on Interaction Design and Children (IDC), ACM Conference on Creativity and Cognition (C\&C), Australian Computer-Human Interaction Conference (OzCHI), ACM SIGCAS/SIGCHI Conference on Computing and Sustainable (COMPASS), ACM Symposium on Eye Tracking Research and Applications (ETRA), Conference on Designing Pleasurable Products and Interfaces (DPPI), ACM Conference on Recommender Systems (RecSys), International Conference on Human-Agent Interaction (HAI), ACM Conference on User Modeling, Adaptation and Personalization (UMAP), International Conference on Mobile Human-Computer Interaction (MobileHCI), ACM SIGACCESS Conference on Computers and Accessibility (ASSETS), and ACM Technical Symposium on Computer Science Education (SIGCSE) \footnote{One paper from SIGCSE appears in the search results and is included in the corpus, as SIGCSE was in cooperation with SIGACCESS in 2008. https://sigcse.org/events/50years.html}. 

Following the same method as the HCI papers, for the Fairness-focused venues, we retrieved 63 papers with ``equit*'' and 15 papers with ``inequit*'' as the search keywords. After removing 10 papers mislabeled as ``Research Article'' and 7 duplicates between the ``equit*'' and the ``inequit*'' results, we retrieved 61 papers.

\subsection{Screening Papers and Finalizing the Dataset } 

We conducted title and abstract screenings to filter out papers irrelevant to equity and technology. Acknowledging that ``equity'' is also a term used in finance denoting the amount of money returned to shareholders, the first and second authors reviewed the title and abstract to filter out papers that use this definition of equity. The two coders divided the corpus in half and independently assessed each paper, assigning one of the codes: (1) equity as defined in the finance context, (2) equity not defined in the finance context, (3) uncertain. The first author then conducted a second review of all the papers in the corpus. The first and second authors discussed the 9 papers that were coded as (3) uncertain. By referring to the full text of these papers, the two authors clarified the uncertainties and concluded that none of these papers define equity in the financial context. Therefore, no papers were excluded in this process. The final corpus included 202 papers, with 141 papers in HCI and 61 papers in the Fairness-focused venues.

\subsection {Data Analysis}

We conducted closed-coding and open-coding to identify trends and surface emerging themes and patterns across our dataset. Our research team met regularly to discuss and organize our analysis, which was recorded into a shared spreadsheet. 

\subsubsection{Open-Coding Analysis} 

To identify themes across the dataset, the research team conducted open-coding centered on motivations, findings, and emerging patterns. Each research member was responsible for open-coding analysis focusing on the paper’s content such as the abstract, title, introduction, methods, findings, and discussions. For the first round of coding, the first author coded 56\% of the corpus, and the second, third, and fourth author each coded approximately 14\%. The first author conducted a second round of coding on the entire corpus, adding additional insights overlooked in the initial round. Throughout this process, the research team met weekly to discuss emerging themes or patterns that highlighted how researchers studied equity. For example, we found that papers often employ a form of stakeholder participation to understand the impacted communities’ underlying needs. Further details of the themes will be discussed in Section~\ref{section5:what}, \ref{section6:why}, and \ref{section7:how_conduct_equity_research}. We did not perform an inter-rater reliability for our open-coding analysis, as the purpose was to iteratively discuss and find key themes that emerged from the corpus, without narrowing the discussion to predefined themes. Existing systematic literature reviews have also used this approach~\cite{sannon2022privacy, disalvo2010mapping}. 

\subsubsection{Closed-Coding Analysis } 
We extracted the descriptive data from all 202 papers in our corpus and organized them in a shared spreadsheet. We organized each paper’s publication venue, publication year, methods, author keywords, list of authors, and the definition of equity if defined in the paper. The methods were categorized into the following five categories: Technical/System, Empirical-Qualitative, Empirical-Quantitative, Design, and Theoretical. The coding was conducted by the first and second authors. The first author cross-checked each code. The closed-coding data was used to understand the overall trends of the HCI and Fairness community such as the number of publications over the years, and statistics about how many papers were published at a certain venue. We further elaborate on these details in Section~\ref{overview}. 

In addition to collecting these descriptive data, we coded each paper according to the interventions and strategies researchers employed to foster equity in technologies. This coding process provided insights on the overall trend in how researchers address equity. Detailed in Section~\ref{section7:how_conduct_equity_research}, our thematic analysis revealed distinct efforts by researchers to integrate equity in technologies. First, researchers monitor and surface existing inequities inherent in technologies and society. Second, researchers develop interventions in three different ways to promote equity in technologies: creating systems and algorithms, developing frameworks, and involving impacted stakeholders in the technology design process. As a result, we created four categories in our coding scheme: ``surfacing and monitoring inequities'' and the three distinct interventions to advance equity. 

We observed that some papers used a combination of two different categories. For example, when Lin et al. addressed health equity, specifically in menstrual health, the authors not only surface adversities to minority groups with minimal menstrual education experience when tracking menstrual data but also conduct a participatory design workshop to design a menstrual tracker addressing the stakeholders’ needs~\cite{lin2022investigating}. For these papers, we specified the two categories the authors used. Moreover, we faced challenges in coding some papers that did not fit in any of the four categories. These papers either briefly mentioned equity to motivate their work or mentioned the term equity without diving deeper into the concept. For instance, Ramesh et al. explored the relationship between on-demand work platforms and workers, specifically understanding how ambiguities of the platform formed the workers' experiences. In this paper, equity is only mentioned in the abstract once~\cite{ramesh2023ludification}. We labeled these papers as ``equity discussed briefly''. The coding for the entire corpus was conducted by the first and second authors, with a moderate inter-rater reliability score (Cohen’s $\kappa$ =0.66)~\cite{mchugh2012interrater}.

 \section{Overview of Dataset}
\label{overview}
In this section, we provide a summary of our dataset, which encompasses the landscape of equity and technology in the HCI and Fairness-focused venues. We discuss the publication venues, the publication year, overlapping authors in HCI and Fairness-focused venues, and the keywords identified by the papers' authors. 

\subsection{Publication Venues and Years}

\begin{table*}[ht]
  \centering
  \caption{Our dataset comprises 16 journals and proceedings in HCI venues and 4 in the Fairness-focused venues. Note that CSCW was a proceedings through 2016 and a journal thereafter. There are 141 papers in HCI venues and 61 papers in Fairness venues in total in our final corpus.}
  \label{tab:dataset-HCI-Fairness}
  
  \begin{tabular}{lp{0.75\textwidth}}
    \toprule
    \textbf{Venues} & \textbf{Journals and Proceedings} \\
    \midrule
    \textbf{HCI (141)} & CHI (56), CSCW (52), DIS (10), GROUP (4), IDC (3), ASSETS (3), C\&C (2), OzCHI (2), COMPASS (2), ETRA (1), DPPI (1), RecSys (1), HAI (1), UMAP (1), MobileHCI (1), SIGCSE (1) \\
    \textbf{Fairness (61)} & FAccT (40), AIES (12), EAAMO (9), SaTML (0) \\
    \bottomrule
    \end{tabular}
\end{table*}

We had a total of 202 papers consisting of 141 in HCI and 61 in the Fairness-focused venues. Illustrated in Table \ref{tab:dataset-HCI-Fairness}, of the 141 papers in HCI category, 56 (39.72\%) were published at CHI and 52 (36.88\%) at CSCW, including papers from both the conference proceedings and the PACM journal. Other HCI venues include DIS (10, 7.09\%), 
GROUP (4, 2.84\%), IDC, ASSETS (3, 2.13\% each), C\&C, OzCHI, COMPASS (2, 1.42\% each), ETRA, DPPI, RecSys, HAI, UMAP, and MobileHCI, and SIGCSE (1, 0.71\% each). In the Fairness category, 40 (65.57\%) were from FAccT, and the others were from AIES (9, 14.75\%), EAAMO (12, 19.67\%) and SaTML (0, 0\%). 

\subsubsection{Growth of Equity in HCI and Fairness} 
\label{equityGrowth}

\begin{figure}[h!]
    \centering
    \includegraphics[width=0.75\textwidth]{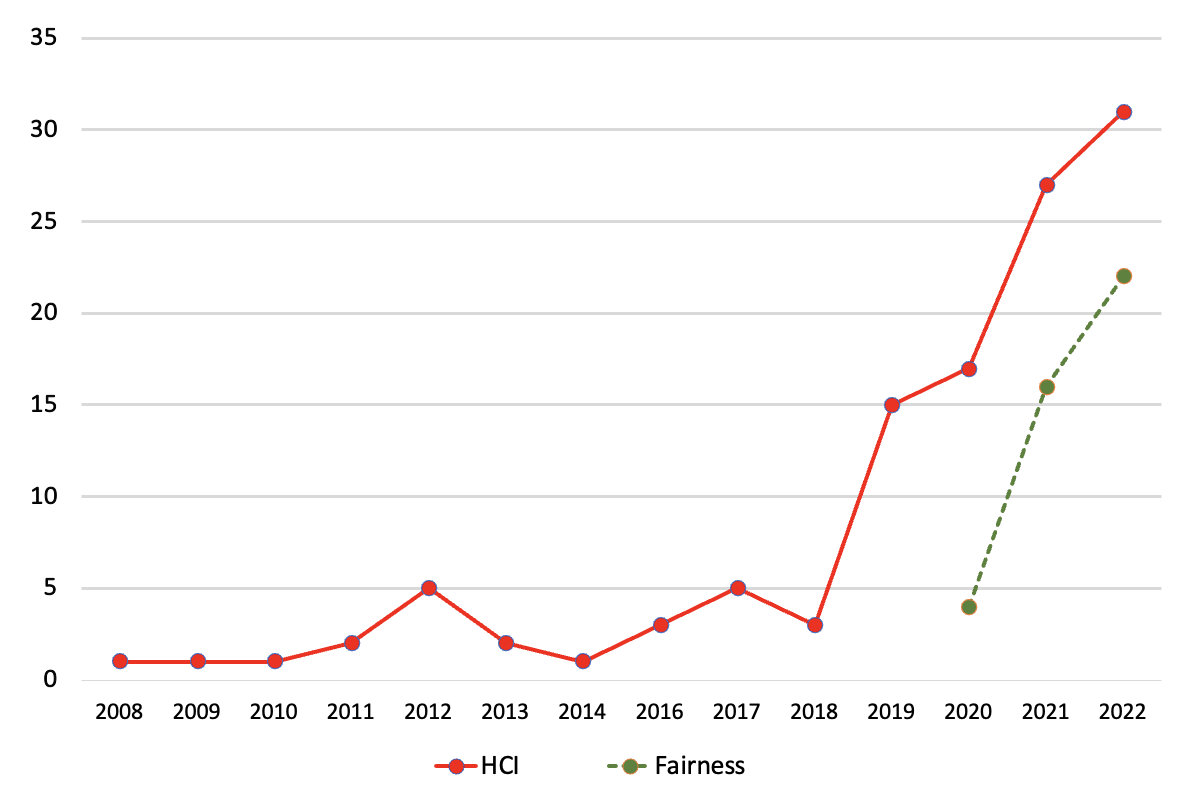}
    \caption{Publication counts in HCI and Fairness community. Our corpus includes papers up to September 2023, but we excluded papers published in 2023 in this figure since all the publications that were to be published in 2023 were not available. There are no data points earlier than 2020 as FAccT joined ACM in 2019 and AIES started publishing papers in 2018. No papers were found prior to 2008 in our corpus. }
    \Description{The trend of the publication year of the research papers underscores an increasing interest in equity and technology across HCI and Fairness in recent years.}
    \label{fig:corpus-year}
\end{figure}

Figure \ref{fig:corpus-year} illustrates the distribution of the papers in HCI and Fairness between 2008 and 2022. We discovered that prior to 2018, there was a dearth of 
equity research with fewer than five papers published annually in both HCI and Fairness-focused venues. Yet, in HCI, the graph illustrates papers spanning from 2008 to 2022, with a notable increase observed from 2018 to 2022. Papers in the Fairness-focused venues were published between 2020 and 2022, reflecting a growth of interest in equity issues corresponding to the increase of HCI papers. This trend underscores an escalating interest in equity and technology in both HCI and Fairness communities in recent years.

It is important to note that while our dataset includes papers published up to September 2023, Figure~\ref{fig:corpus-year} does not cover papers from 2023, as the data for the full year was not available. Unlike other systematic literature reviews that restrict their dataset to the most recent 10 years~\cite{sannon2022privacy}, we did not impose restrictions on the timeframe of when the papers were published for our final dataset. Considering the first published paper to discuss equity was in ACM-DL SIGACCESS in 2008, we considered this year as fairly recent.

\subsection{Authors, Author Keywords, and Context} 
\label{section4:authors-keywords-context}
We found 538 unique scholars in HCI and 311 scholars in Fairness-focused venues. To understand how many scholars reside in the intersection of these two communities, we analyzed the number of researchers who published at both HCI and Fairness-focused venues within our corpus. We found 6 researchers (0.7\% of all unique scholars in our corpus) in our corpus published to both HCI and Fairness-focused venues within our corpus. Additionally, we analyzed the author keywords to better understand the context in which equity is discussed. Author keywords are terms identified by the authors to demonstrate the overall theme and context of the research. Details of the overlapping authors are illustrated in Appendix~\ref{appendix:authors}.

Illustrated in Appendix~\ref{appendix:author-keywords}, the most frequently used keywords in HCI were “equity”, “participatory design” and “gender”, followed by “accessibility” and “feminist hci”. The frequency of these keywords indicates that the papers in our corpus were commonly focused on gender and accessibility issues. In the Fairness-focused venues, “fairness” and “algorithmic fairness” were used most frequently, indicating that discussions of equity are often framed as “fairness”. 

\section{What Shapes Equity Research?}
\label{section5:what}

To understand the factors shaping equity research, we elaborate on the definitions researchers use to define equity, the lenses and frameworks they draw on when discussing equity, and the methodologies they use when studying equity.

\subsection {What Definitions (if any) Do Researchers Use for Equity?}
Providing an explicit definition of ``equity'' is uncommon, with only 27\% (Total=54) in the entire corpus with a clear definition. Specifically, 21\% of the HCI papers (HCI=30) defined equity and 40\% of the Fairness papers (Fairness=24) had a explicit definition of equity. These papers frequently mention ``equity'' without a definition or use the term interchangeably with related terms such as inclusivity, accessibility or fairness. For example, Varanasi et al. discussed the challenges of Machine Learning practitioners executing responsible AI values such as fairness, transparency and accountability in AI/ML based products. The authors did not provide a definition of equity but mentioned equity as one of many responsible AI values~\cite{varanasi2023it}. However, in papers where the definition of equity is explicit, there is no universal agreement on what equity means. Some definitions of equity involve treating people differently (that we identified as need-based and contribution-based definitions). On the other hand, others defined an equitable practice as treating people the same (equality-based definition). Below, we describe and discuss these definitions.

\subsubsection{Need-Based Definition} 
Need-based definition means prioritizing the distribution of resources or attention based on the \textit{needs} of particular stakeholders~\cite{solyst2023would, lee2019webuildai, quinonerocandela2023disentangling, mota2020desiderata}. For example, Lee et al. created a participatory framework where impacted communities collectively build their own algorithmic policies. In the context of allocating food donation transportation services, the authors defined equity-based allocation as ``giving donations to recipients with greater need''~\cite{lee2019webuildai}. 

Quiñonero-Candela et al. introduced an AI fairness framework at LinkedIn to unravel how to define fairness in AI systems that decide job candidate recommendations to recruiters, suggest connections or recommend job postings. In this paper, the authors explicitly differentiate the definition between equity and equality. While equality means ``treating every candidate the same'', equity is defined as ``investigat[ing] whether there has been historical barriers [...] or whether the qualifications required are too narrow and would exclude females capable of performing the job as well as males''~\cite{quinonerocandela2023disentangling}. Hence, unlike equality, according to Quiñonero-Candela et al. equity takes into account the candidates' backgrounds and needs when determining their qualifications for the job~\cite{quinonerocandela2023disentangling}. Moreover, Mota et al. specified equity as ``need-based principles of allocating resources''~\cite{mota2020desiderata} in the context of donating money to public schools. An equitable donation involves allocating money to public schools that often fail to receive sufficient funding~\cite{mota2020desiderata}.

\subsubsection{Contribution-Based Definition} 
Contribution-based definition is relevant to equity theory, which evaluates fairness by comparing the ratio between one's input and outcomes with that of others. Inputs can be defined as an individual’s contributions to an organization, while outcomes are the corresponding rewards one receives~\cite{adams1963towards}. Employing equity theory, Uhde et al. discussed equity within the context of collaborative shift scheduling systems~\cite{uhde2020fairness}. D’eon et al. incorporated equity theory by defining equitable treatment as ``rewards should be proportional to the quality or quantity or the time they spend on the job''~\cite{deon2019paying}. 

\subsubsection{Equality-Based Definition}

When some papers discuss equity, they mean equality. We observed this pattern particularly when researchers discuss equity in the context of measuring engagement levels of individuals in a group setting. For instance, ``Equity of participation'' or ``verbal equity score'' is often used to determine whether there was a balanced verbal or non-verbal communication within a group~\cite{borge2012patterns}. While a low equity of participation indicates that a single person dominates the group activity, a high equity of participation represents harmonized engagement levels among the group members. Researchers measure equity of participation through Gini Coefficients, a scale ranging from 0 to 1 quantifying the variance in engagement levels among group members. A value closer to 0 indicates a higher equity of participation and a balanced contribution among group members~\cite{basheri2012multitouch, westendorf2017understanding, wallace2013collaborative}. 

Aside from measuring engagement levels, papers often employed equality-based metrics such as demographic parity and other forms of group-based statistical parity to measure equity~\cite{reader2022models, halevy2021mitigating, halevy2021mitigating, chohlas-wood2021blind, fabris2022tackling}. For example, Reader et al. defined equity as ``equal statistics across groups'' in the context of developing a feedback system in social systems to address societal inequities~\cite{reader2022models}. Hsu et al. proposed a US Census-linked API that conducts an equity analysis measuring how resources are allocated by the local government. The authors used demographic parity-based fairness which ``test(s) for statistically significant differences in response time across different groups of zip codes''~\cite{hsu2021open}. Fabris et al. conducted a survey on datasets used in research for algorithmic fairness. The researchers discussed the operationalization of equity as ``measuring disparity in some algorithmic property across individuals or groups of individuals''~\cite{fabris2022tackling}.

\subsection {What Lenses and Frameworks Do Researchers Draw on When Studying Equity?}
\label{section5_lenses_framework}
To delve into equity, several authors employed equity-relevant lenses and frameworks as the central focus of their research. Notably, equity-relevant concepts including a) Feminist HCI, b) social justice, and c) ableism were discussed either as primary lenses of the paper or as frameworks to explore the design implications derived from the researchers’ findings. 

Employing and discussing feminism is one of the most common frameworks used by researchers in discussing and studying equity and technology~\cite{bardzell2010feminist, madden2021why, menking2019people, petterson2023playing, kaur2023oh, strengers2020adhering, harrington2019deconstructing, coenraad2019enacting, wagman2021command, hope2019hackathons, su2022what, mccradden2023what, katell2020situated}. To systematically integrate feminism to HCI, Feminist HCI~\cite{bardzell2010feminist} offers key principles: \textit{pluralism, participation, advocacy, ecology, embodiment, self-disclosure}. For example, pluralism argues that human experiences are too varied to narrow down to one universal solution. Researchers apply these Feminist HCI principles to explore design implications to create equitable outcomes. Madden et al. suggested leveraging Feminist HCI principles to create an inclusive and participatory esport culture~\cite{madden2021why}. Based on discriminatory personal experiences identified by professional gamers and event organizers, Madden et al. suggested the importance of \textit{advocacy} that challenges institutionalized gender biases in the design of esports~\cite{madden2021why}. Wagman et al. challenged the assumption that machines are simply tools used by humans. The authors employed feminist science and technology studies (STS) to argue for an equitable relationship between humans and machines. Creating this just and inclusive relationship challenges the assumption that machines lack agency and are politically neutral~\cite{wagman2021command}. 

Along with feminism and Feminist HCI, social justice plays a crucial role when discussing equity \cite{tseng2020we, desportes2021examining, vecchione2021algorithmic, petterson2023playing}. Petterson et al. utilized Nancy Fraser’s dimensions of social justice \cite{fraser2008social} to address the need to redistribute the power and recognize community perspectives when exploring design toolkits that aim to promote equity~\cite{petterson2023playing}. Similarly, Desportes et al. applied Dombrowski et al.'s social justice framework~\cite{dombrowski2016social} to investigate how makerspaces could be positioned to surface social inequities~\cite{desportes2021examining}. In addition to social justice, researchers discussed ableism when designing tools too address accessibility~\cite{shinohara2021burden}. Treating nondisabled people as superior, ableisem as a lens enables researchers to critically examine assumptions integrated in design. Shinohara et al. leveraged ableism to identify inaccessibility experienced by graduate students and proposed equitable solutions that addresses accessibility gaps in a graduate school setting~\cite{shinohara2021burden}. 

\subsection{Overview of Methodologies}
\label{methodologies}
\begin{figure}[h]
    \centering
    \includegraphics[width=0.75\textwidth]{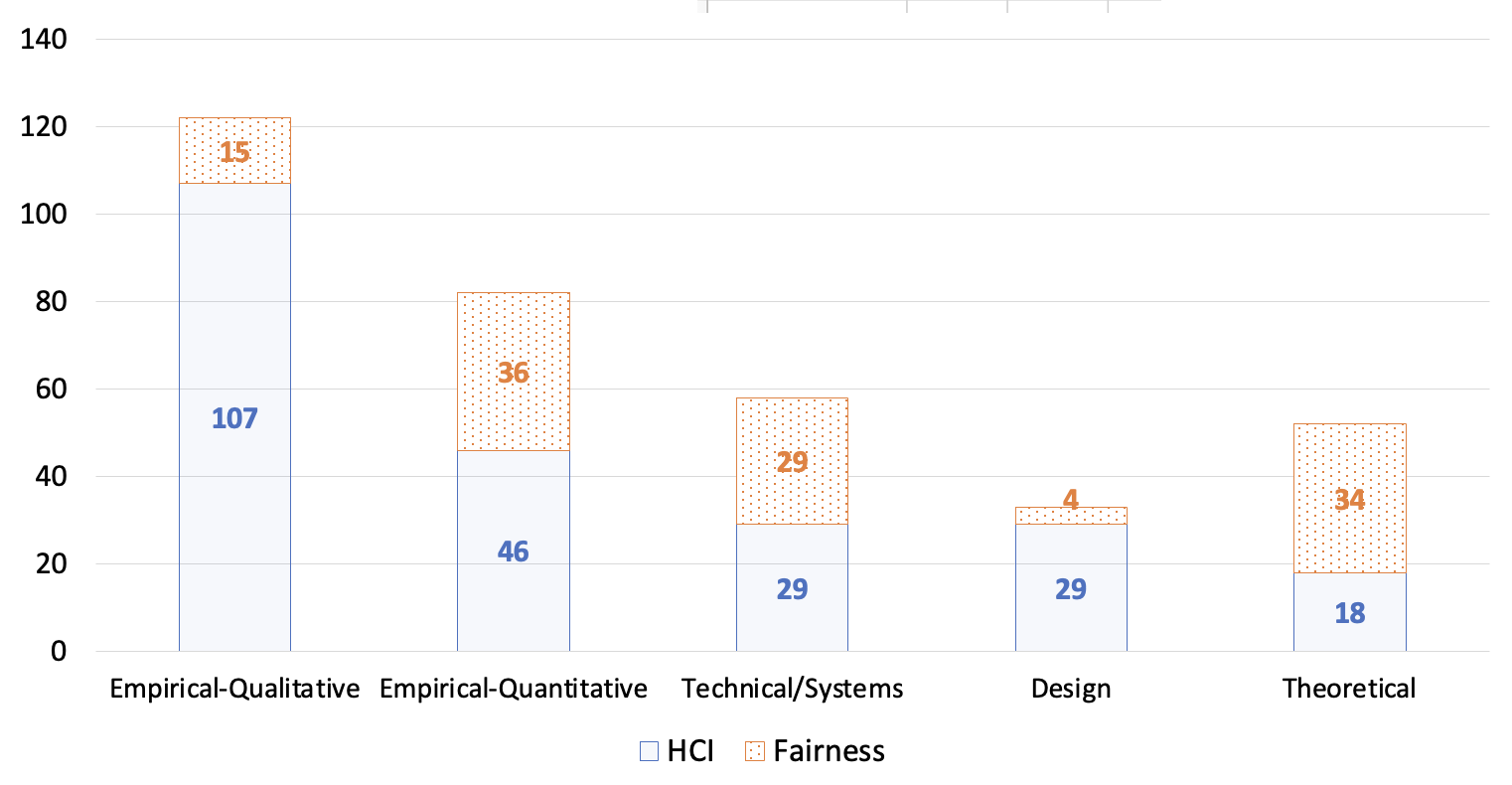}
    \caption{Methods distribution for HCI and Fairness papers. We counted the methods used for each paper. Papers that use multiple methods are counted more than once.}
    \Description{The visualization of the distribution of six research methods (Empirical-Qualitative, Empirical-Quantitative, Technical/Systems, Design, Theoretical) used in HCI and Fairness papers.}
    \label{fig:methods-value}
\end{figure}

In this section, we provide a detailed illustration of methodologies used in HCI and Fairness. We categorized all papers into six categories: Empirical-Qualitative (e.g.,~\cite{su2022what, houtti2023all, puussaar2022sensemystreet}), Empirical-Quantitative (e.g.,~\cite{adam2022write}), Technical/Systems (e.g.,~\cite{krafft2021actionoriented, xie2022surfacing, marchiorimanerba2022investigating}), Design (e.g.,~\cite{coenraad2019enacting, katell2020situated, sakaguchi-tang2021codesign}), and Theoretical (e.g.,~\cite{asad2019prefigurative, quinonerocandela2023disentangling}). As shown in Figure~\ref{fig:methods-value}, the distribution for each method is as follows: Empirical-Qualitative (HCI = 107, Fairness =15), Empirical-Quantitative (HCI = 46, Fairness= 36), Technical/Systems (HCI = 29, Fairness=29), Design (HCI = 29, Fairness=4), and Theoretical (HCI = 18, Fairness= 34). We counted the methods used for each paper; therefore, papers employing multiple methods are counted multiple times. Most papers in the HCI used Qualitative methods (76\%), whereas the Fairness community primarily used Quantitative (59\%) and Theoretical (56\%) methods.

While labeling each papers’ methodologies, we found some papers used more than one method (HCI = 67, Fairness = 42). In HCI, the most commonly used combinations of methods were Qualitative and Quantitative (N=15) (e.g.,~\cite{haimson2021disproportionate,foong2021understanding,wallace2013collaborative}) as well as Design and Qualitative (N=15) (e.g.,~\cite{harrington2019deconstructing, stein2023you, higgins2023investigating}). In Fairness, Technical/Systems and Quantitative was the most used combination (N=15) (e.g.,~\cite{bashardoust2023reducing, puranik2022dynamic, dejonge2023unfair}). See Appendix~\ref{appendix:methods-details} for further details on the mixed-methods used in the dataset. In the following sections, we elaborate how understanding the researchers' methodological approaches align with researchers' efforts to integrate equity in technology. 

\section{Why Study Equity and Technology?}
\label{section6:why}
In this section, we elaborate why researchers focus on equity and technology. We found that their motivations to study equity were derived from addressing existing inequities in technology and society, as well as the research community's lack of focusing equity as an explicit goal. 

\subsection {Existing Inequities in Society and Technology}
Researchers often discuss existing inequities that are amplified by technologies or societal issues as motivations of their work. These motivations are often introduced in the beginning of the paper when introducing the study's context. Several studies have discussed that existing technologies such as algorithmic decision-making systems inherit, codify, and amplify existing societal inequities, particularly highlighting discrimination and challenges faced by marginalized communities such as immigrants~\cite{liaqat2020leveraging}, transgender individuals~\cite{haimson2020designing} or Black Americans~\cite{martin-hammond2022bridging,brewer2023envisioning}. For example, Haimson et al. illustrated not only the challenges trans people experience but also the limitations of current technologies in supporting this population. Trans individuals face unique barriers including discrimination, violence and lack of resources to support their well-being. Since technologies are not designed to support these challenges, Haimson et al. argued for technologies that meet trans individuals' needs~\cite{haimson2020designing}. 

Researchers aimed to address “digital divide”, often defined as inequitable access to digital technologies which magnify existing inequities. For example, Du et al. highlighted the digital divide among older adults, especially those with lower socioeconomic status and those who lack not only the access to technologies but also the skills to use technologies. To address this inequity, Du et al. collaborated with low-income senior housing organizations, instructing seniors on how to use tablet PCs~\cite{du2021lessons}.

Another context that researchers explored is health inequities. For instance, Karusala et al. illustrated that the inequitable access to healthcare systems in the Global South motivated the researchers to develop a patient education program focused on maternal and child health~\cite{karusala2023unsettling}. Amugongo et al. emphasized the importance of developing ethical guidelines and principles for healthcare AI systems to address ethnic minorities' needs. The authors argued that healthcare AI systems not only exacerbate existing inequalities but also have been built upon Western values~\cite{Amugongo2023invigorating}. Similarly, Martin-Hammond et al. emphasized that Black communities are disproportionately affected by cardiovascular diseases. Black Americans face challenges in meeting the dietary and physical activity requirements for a healthy heart due to inequities in accessing proper care. Hence, the authors investigated how Black Americans engage with behavior change programs that motivate behaviors for a healthy heart and elaborate design implications for community-based personal informatics tools that is culturally informed and addresses Black communities' needs~\cite{martin-hammond2022bridging}.    

\subsection {Failures to Address Equity Goals }
\label{section6:6.2_failures}
We found that researchers not only address the limitations of existing technologies but also highlight the insufficient efforts within the research community to achieve equity goals. Researchers highlighted that initiatives to merely mitigate bias are not enough to create equitable outcomes. For example, Solyst et al. illustrated that culturally responsive STEM and computing initiatives focus on the learners’ identities and experiences. With the increase of online learning and virtual educational experiences, the authors addressed that researchers have developed technologies that support positive virtual social interactions. Yet, Solyst et al. emphasized that these technologies and research efforts do not sufficiently address the learners’ cultural contexts, lacking efforts to develop equitable outcomes~\cite{solyst2023would}. Similarly, Xie et al. emphasized that to accomplish equitable teaching, receiving students’ feedback is crucial for addressing minority students’ unique struggles~\cite{xie2022surfacing}. Prior research focused on examining online course systems that afford instructors to better understand students’ performances, and experiences in the course. However, Xie et al. addressed that these systems fall short of contextualizing student feedback with the students’ lived experiences, which is crucial for promoting equity~\cite{xie2022surfacing}. 

Additionally, researchers explored limitations of existing language and speech technologies that focus on mitigating biases rather than addressing equity goals.
For instance, Strengers et al. discussed the limitations of existing endeavors by natural language processing researchers in addressing gender biases exacerbated by technologies. As means to debias these natural language technologies, removing or hiding any sensitive attributes that indicate gendered information has been a common solution. Yet, this solution simply neutralizes or removes gendered languages altogether rather than seeking to address systematic gender stereotypes or supporting minorities' interests~\cite{strengers2020adhering}. Similarly, Papakyriakopoulos et al. emphasized that the training datasets used for speech-language technologies lack diversity in language dimensions such as accent, dialect, speech impairment and intersectionality of demographic features. For example speech-based hiring screening software tools may discriminate against applicants who do not speak the "standard" language. Therefore, the authors curated an augmented datasheets for speech datasets with the goal to support practitioners and researchers to create ethical datasets that lead to equitable and inclusive speech language technologies~\cite{papakyriakopoulos2023augmented}.

\section{How Do Researchers Conduct Equity Research?}
\label{section7:how_conduct_equity_research}
While Section~\ref{section6:why} discusses why researchers focus on equity and technology, in this section, we address how researchers conduct equity research through different approaches when advancing equity (Table~\ref{tab:focus}). We found a group of papers in our corpus marginally focusing on equity, and using equity merely as a motivation for their research, without a focus on studying or advancing equity. The rest of the papers which focus on equity either invest in monitoring and surfacing inequities in existing technologies, and/or developing interventions to foster and integrate equity into technology design and development. These interventions include a) developing systems/algorithms to provide more equitable opportunities for underserved and marginalized stakeholders, b) framework-based interventions, and c) employing stakeholder participation processes towards a more equitable technology design. We elaborate how the authors' methodological choices align with how the authors decide to address equity. When developing these interventions, researchers addressed several tensions and trade-offs between equity and other values such as utility-based (e.g. efficiency and accuracy), social values (e.g. equality, privacy), and even equity itself.

\begin{table}[h]
    \centering
    \caption{Distribution of papers based on whether the paper uses equity merely as a motivation, monitors and surfaces inequities, and/or develop interventions to foster and promote equity in technology design and development. The percentage represents the number from each respective community, and the Total column is the percentage of the entire corpus. Papers may fall into multiple categories.}
    \label{tab:focus}
     \begin{tabular}{p{0.36\textwidth} p{0.16\textwidth} p{0.16\textwidth} p{0.16\textwidth}}
        \toprule
        & \textbf{HCI} & \textbf{Fairness} & \textbf{Total} \\ 
        \midrule
        \textbf{Equity Merely as Motivation} & 21\% (n=29) & 21\% (n=13) & 21\% (n=42) \\
        \midrule
        \textbf{Monitoring and Surfacing \newline Inequities} & 47\% (n=66) & 10\% (n=6) & 36\% (n=72) \\
        \midrule
        \textbf{Intervention} & 43\% (n=61) & 84\% (n=51) & 55\% (n=112) \\ 
        \quad System/Algorithm & 16\% & 39\% & 23\% \\ 
        \quad Framework & 8\% & 38\% & 17\% \\ 
        \quad Stakeholder participation & 19\% & 7\% & 15\% \\ 
        \bottomrule
    \end{tabular}
\end{table}

\subsection{Equity Merely as a Motivation}
21\% of papers in our dataset only marginally focus on equity. These papers mention “equity” as a motivation for conducting the research without expanding on the concept or developing equity-based interventions. As discussed in Section~\ref{method}, we labeled these papers as “equity discussed briefly”. We identified a total of 42 papers that briefly discuss equity in our corpus: 29 papers (21\%) from HCI and 13 papers (21\%) from the Fairness-focused venues. These papers mention the word “equity” or “inequity” to primarily set the stage for their research and provide background information. However, the main research questions of these papers do not focus on equity but rather on relevant concepts such as fairness or privacy. For example, Van Berkel et al. discussed the need to identify fair predictors to create equitable algorithms in recidivism. However, the authors focused on participants’ perceptions of fairness rather than delving into equity~\cite{van2019crowdsourcing}. In another work, Pfohl et al. incorporated fairness considerations in healthcare algorithms motivated to mitigate exacerbation of health inequities. However, the authors did not develop equity-focused interventions or metrics to assess equity~\cite{pfohl2022net}. From these examples, we observed how equity is discussed superficially among researchers without further details or definitions. We suspect this pattern may stem from the research community’s lack of awareness of the equity definitions researchers have discussed in Section~\ref{section5:what}. 

\subsection{Monitoring and Surfacing Biases and Inequities }
Researchers have focused on identifying inequities stemming from current technologies and understanding the impacts these technologies have on affected stakeholders, particularly underserved and marginalized communities. This focus on identifying existing inequities in technologies was more prevalent in the HCI literature (47\%, HCI=66), than the fairness literature (10\%, Fairness=6). Some of these identified inequities include : gender \cite{menking2019people, murphy-hill2023systemic, madden2021why, im2022women, offenwanger2021diagnosing}, education \cite{wani2022hartal}, race \cite{offenwanger2021diagnosing, houtti2023all}, disabilities \cite{shinohara2021burden}, and workplace inequities \cite{houtti2023all, khatri2022social, brown2022attrition}. To identify such inequities, researchers often employed qualitative and quantitative methods, as means to collect first-hand experiences. 
 
Qualitative methods include conducting interviews with impacted stakeholders to surface challenges and inequitable experiences. Menking et al. conducted a interview-based study with women Wikipedians to understand how they perceive both physical and conceptual safety on Wikipedia. Highlighting the burden that women Wikipedians undergo to ensure their safety both on and off the platform, Menking et al. illustrated how their findings align with the experiences of marginalized groups in online communities~\cite{menking2019people}. Murphy-Hill et al. described gender inequities in the code review processes at Google, where the recommendation system for distributing code reviews exacerbates human biases. The recommendation system has created disparities in the number of reviews received between men and women~\cite{murphy-hill2023systemic}. Wani et al. illustrated the cultural context behind 
Kashmire, India’s technology and education infrastructure, delving into how socio-political conflicts disrupt children's opportunities to continue their educations~\cite{wani2022hartal}. Jin et al. surfaced digital inequities highlighting the challenges and coping strategies for older adults in China when utilizing digital banking platforms~\cite{jin2021too}. Shinohara et al. identified existing inequities for Ph.D. students who are identified as blind or low vision, or deaf or hard to hear. The authors investigated the insufficient accommodations for these students~\cite{shinohara2020access}. 

Research also employed quantitative analysis to monitor inequitable practices. Toxtli et al. created a plugin to monitor equity by quantifying invisible labor in crowd work, particularly in Amazon Mechanical Turk. Measuring the amount of invisible labor hours ensures workers are paid equitably for their work. This tool’s effectiveness was evaluated through participant surveys and accuracy of detecting the number of invisible hours~\cite{toxtli2021quantifying}. Additionally, Breen et al. conducted a large-scale national survey to measure cybercrimes in the United States. Researchers gathered empirical evidences of inequities within cybercrime victimization. Demographic analysis showed that older Americans and Black Americans are more likely to be victimized. Based on this finding, Breen et al. suggested the need for researchers to further investigate cybercrimes within the context of marginalized populations~\cite{breen2022largescale}.

\subsection {Interventions to Foster Equity}
We observed three lines of effort to advance equity in technology design and development: a) developing systems/algorithms that provide equitable opportunities for underserved and marginalized stakeholders, b) introducing frameworks for equitable outcomes and c) employing stakeholder participation processes towards an equitable technology design. Few researchers evaluated their interventions to determine the efficacy or effectiveness of proposed tools/models or frameworks in improving equity. The prevalence of such interventions was higher in the Fairness community (84\%) compared to the HCI community (43\%). We observed that the Fairness-focused venues prioritize integrating or assessing equity in diverse contexts such as public transportation~\cite{tedjopurnomo2022equitable}, recruitment systems~\cite{quinonerocandela2023disentangling}, criminal justice~\cite{chohlas-wood2021blind}, and IRS tax systems~\cite {black2022algorithmic}. In this section, we provide details on the interventions developed by both the HCI and Fairness communities.

\subsubsection{System/Algorithm Development Interventions to Provide More Equitable Opportunities for Underserved and Marginalized Stakeholders} 

We observed a series of interventions through the development of algorithms and systems specifically designed to promote equity for impacted stakeholders, particularly those in underserved and marginalized communities. These interventions encompass a wide range of approaches: a) designing and creating accessible and inclusive systems for impacted stakeholders, b) encouraging and empowering other stakeholders to support the impacted communities, c) and algorithmically promoting equity in decision-making processes. We saw a trend where the Fairness community (39\%) had a higher percentage of developing system/algorithm development interventions than that of HCI (16\%). This trend also aligns with Fig~\ref{fig:methods-value}, which shows that a high percentage of the Fairness-focused venues use technical/systems and quantitative methods. Below, we describe some of these efforts.

One set of interventions focused on designing and creating inclusive systems for underserved and marginalized populations. For example, Uchidiuno et al. addressed educational equity in rural and underserved communities by developing a K-3 learning system that curates an equitable help-seeking culture. The system pairs students as helpers and tutees with equal probability, giving low-achieving students the opportunities to become helpers in areas they are proficient in. Using mixed-methods approach, the authors measured not only the students' interactions during the group sessions but also the engagement levels and the number of activities the students mastered in the learning system. The findings suggested that the intervention fostered interactions among peers and resulted in students selecting challenging tasks in the education system~\cite{odiliuchidiuno2021fostering}.

In another line of work, researches developed systems to encourage and empower various stakeholders to make equitable decisions and support impacted communities. For example, Mota et al. created design interventions in an online donation platform to nudge donors to make equitable donations to online education charities, ensuring that impoverished schools receive more donations. The key design consideration was to guarantee the donors' freedom of choice. The authors developed two design interventions: an explicit design that openly shows the poverty level of each school listed on the platform and an implicit design that, by default, ranks the schools based on the schools’ economic needs. Conducting an experimental study, Mota et al. found that nudging users through these interventions resulted in higher donations to schools with the highest poverty level. The authors concluded that nudging users through interface design could help achieve equity goals~\cite{mota2020desiderata}.   

Another group of work focused on algorithmically promoting equity in decision-making processes. For example, to address equitable decision-making in the criminal justice system, Chohlas-Wood et al. created an algorithm that redacts any racial cues in criminal incident reports used to predict a defendant’s recidivism. Amidst controversies that the criminal justice system discriminates against Black defendants, Chohlas-Wood et al. developed an algorithm that hides race-related information including proxies for these features without obscuring the narrative of felony incident reports. The algorithm was evaluated by assessing whether prosecutors made the same charging decisions between race-redacted felony cases and the original document that includes protected attributes. While the findings indicated that the system did not significantly change the charging rates, the system provided an efficient way of redacting racial cues that would otherwise require humans to do the job~\cite{chohlas-wood2021blind}. 

\subsubsection{Framework-based Interventions}  
In our analysis, we identified interventions aimed at advancing equity by developing frameworks~\cite{bynum2021disaggregated, nee2021advancing, hsu2021open, quinonerocandela2023disentangling, lee2019webuildai, strengers2020adhering}. Carroll defines frameworks as “design guidelines based on perception which are assumed to be relatively permanent and are often accepted across different cultural contexts”~\cite{carroll2003hci}. Using this definition of framework, we found researchers have developed frameworks to provide an initial starting point for the research community to investigate and inspire further research on the topic. Frameworks can be adapted to different case studies beyond what the author originally presented~\cite{ nee2021advancing, hsu2021open}. In our corpus, 8\% of the HCI community and 38\% of the Fairness-focused venues were framework-based interventions.

Nee et al. developed a Natural Language Processing (NLP) framework that advances linguistic justice and prevents NLP tools from exacerbating biases embedded in human language. This framework is composed of four layers of linguistic structures (words and phrases, organization of words and phrases, patterns of language use over time, power inequities). To evaluate the framework, Nee et al. conducted a case study at a large Silicon Valley tech firm that aimed to develop NLP tools that automatically detect harmful phrases and recommend non-discriminatory terms~\cite{nee2021advancing}. 

Hsu et al. developed a framework aimed at helping local government agencies address inequities by curating a systematic and standardized analysis of government data. The framework standardizes data collection, including demographic information, and incorporates analytical approaches that assess inequity. This framework is composed of three components: 1) a US census-linked API which incorporates demographic information to identify historically underserved communities; 2) an equity analysis playbook, which quantitatively summarizes how government services and resources are allocated 3) a standardized method for government datasets. Hsu et al. demonstrated a case study applying the framework to the San Jose fire responses in 2020 and 311 service responses. Through this case study, the framework provided actionable insights for local governments and identified opportunities for equitable distribution of resources~\cite{hsu2021open}

\subsubsection {Employing Stakeholder Participation Processes towards a More Equitable Technology Design}
Involving diverse stakeholders in the process of the design and development of technologies is a common approach to foster equitable technology, as shown by 19\% of the HCI community, and 7\% of the Fairness community, constituting 15\% of the entire corpus. This approach includes various stakeholder participation methods such as participatory design (PD), co-design, value-sensitive design, and participatory action research (PAR)~\cite{delgado2021stakeholder}. These methods surface stakeholders’ needs, promote stakeholder inclusivity, and create a contextual understanding of stakeholders’ challenges, thereby fostering equity in the design of technologies. Among these methods, participatory design (PD) emerged as one of the most frequently used design methods for identifying stakeholders’ needs~\cite{stein2023you}. For example, PD was employed to design and reimagine healthcare support systems, particularly with marginalized communities~\cite{hope2019hackathons,lin2022investigating, stowell2020investigating}. Hope et al. used PD to reimagine support systems for breastfeeding by centering marginalized voices, especially LGBTQ+ individuals and families~\cite{hope2019hackathons}. Similarly, Lin et al. aimed to understand attitudes and perceptions of menstrual cycle tracking by those who have limited resources to menstrual health care~\cite{lin2022investigating}.  

Similar to PD, participatory action research (PAR) focuses on partnering with affected communities and collaborating with them in research plans and developing interventions~\cite{delgado2021stakeholder, katell2020situated}. For example, Katell et al. partnered with American Civil Liberties Union of Washington to surface the organizations’ priorities, solicit feedback, develop prototypes and address issues of fairness and equity in technology policies. As a result of partnering with the organization, Katell et al. proactively applied the organization's feedback suggesting that researchers engage with historically marginalized communities who are underrepresented in surveillance policy decision-making~\cite{katell2020situated}.   

Researchers broadly mention co-design as a terminology to indicate collaborating and developing design interactions with stakeholders. Cruz et al. conducted a co-design study through storyboards with marginalized communities to discuss the design and use of wearable technologies. Although wearable technologies have the potential to benefit the health and safety of underserved populations the most, marginalized communities are often not involved in the design process. Hence, the co-design study illustrating diverse concepts and use cases, the participants voiced how these technologies could be incorporated in their own lives~\cite{cruz2023equityware}. 

We also saw cases of researchers incorporating value-sensitive design (VSD) to surface stakeholders' values and address value conflicts, particularly those of underrepresented stakeholders. 
For example, Robertson et al. conducted VSD in the context of an algorithm that assigns students to public schools. Through VSD, the authors highlighted a mismatch between the values of the district that guided the development of the algorithm and those of the parents. The algorithm failed to address the impacted stakeholders' needs~\cite{robertson2021modeling}.
Other examples include using VSD in designing an inclusive gallery experience for people with visual impairments by surfacing the values of diverse stakeholders including gallery staff, artists, and those with visual impairments~\cite{holloway2019making}.

\subsubsection{When Stakeholder Participation is Not Equitable Itself}
While the previous section highlights how researchers utilize stakeholder participation and cater to their needs in promoting equity, our review also found papers critiquing this approach. These papers argued that enhancing stakeholder participation alone may be insufficient to ensure a truly equitable design process~\cite{sloane2022participation}.  
Merely involving stakeholders in the design process does not automatically address the fundamental challenges of inequities. There is a critical need to tackle potential exploitative and extractive methods while engaging with communities~\cite{sloane2022participation}.
Harrington et al., for example, critiqued that PD-an approach originally intended to democratize design processes-is a privileged activity overlooking the challenges underserved populations experience when asked to envision equitable design solutions. To achieve equitable PD, it is crucial to comprehend the research setting’s historical context, relationship with the community, and recognize unintentional harms created during the design process. As historically marginalized participants tend to distrust the researchers’ intentions for collaborations, building rapport with participants is important. This can be initiated through a formal introduction emphasizing the research team’s desires to learn from the participants; thus, shifting the power dynamic towards the participants. Moreover, another way is to connect with community leaders to gain contextual background of the participants' experiences~\cite{harrington2019deconstructing}. 

Similarly, Prost et al. explored a walking-based participatory method that advocates for equitable researchers-participant relationships, providing participants with greater authorities to lead the discussions and walking routes during the study~\cite{prost2023walking}. Sakaguchi-Tang et al. proposed an equitable design process between student designers and older adults by suggesting sharing life experiences for relationship building and engaging older adults in the study planning. This process overcame age-related barriers and fostered a balanced engagement during the co-design sessions~\cite{sakaguchi-tang2021codesign}. Moreover, Bray et al. emphasized the lack of representation in Black communities, specifically in speculative design practices. Inspired by Afrofuturism, the authors developed method cards as design probes to foster critical design thinking~\cite{bray2021speculative}. Uchidiuno et al. highlighted the importance of understanding the cultural and socioeconomic backgrounds when conducting a co-design study with after school programs. The authors suggested researchers to partner with afterschool center administrators early in the process of developing co-design materials. This approach not only ensures that the co-design study attuned towards the context and culture of the school but also fosters a sense of ownership for the administrators~\cite{uchidiuno2023little}.

\section {Tensions and Trade-offs in Advancing Equity}
\label{section8:tensions}
Researchers encountered various tensions and trade-offs during the process of advancing equity. These tensions and trade-offs happen when a conflict between equity and other values arises, including equity itself.
We identified and categorized these tensions and trade-offs into three groups: 1) equity versus utility-based values 2) equity versus social values 3) equity versus equity values. 

\subsection{Equity versus Utility-based Values} 
Researchers have discussed the need to navigate trade-offs between equity and values such as efficiency and accuracy. 
These trade-offs were evaluated through numerical metrics of equity and utility-based values of a given system. For example, Tedjopurnomo et al. discussed the trade-off between efficiency and equity when optimizing a public bus network. An equitable bus route may reduce efficiency by adding bus services to disadvantaged areas, lowering the directness of the bus routes. On the other hand, an efficiency-based bus route would neglect providing services to disadvantaged areas. The researchers defined equity and efficiency metrics that were applied to three different bus network optimization models. These three models include a direct method that connects two bus stops using the shortest road distance, a gradual method that connects two bus stops by gradually adding bus stops, and a gradual-equity method that considers equity scores allowing detours in the bus routes. In measuring the trade-offs between the equity and efficiency scores for each method, the gradual-equity method sacrificed efficiency to add services in disadvantaged areas and the gradual method failed to detect nearby bus stops that needed to be covered. Researchers concluded that developing a universal solution that solves the trade-offs between equity and efficiency is challenging and may not capture the contextual information of the city~\cite{tedjopurnomo2022equitable}. 

Similarly, Black et al. discussed the trade-offs between accuracy and vertical equity in IRS tax audit models, which monitor individuals who are prone to misreport their tax liability. Vertical equity ensures that tax enforcement is fairly allocated differently across different income levels. Although misreported tax liability is higher for those with the highest income, lower-to-middle income earners are the ones subjected to more frequent audits. Hence, the authors explored how modern machine learning models impact vertical equity. The authors found that improving the model's flexibility increases its accuracy; however, it exacerbates inequities by disproportionately focusing the audit burden to lower-to-middle income taxpayers~\cite{black2022algorithmic}.

Previous examples illustrate researchers navigating trade-offs between equity and utility-based values, often assuming that these values are in conflict with one another. However, Niu et al. addressed a scenario where equity and accuracy are not necessarily always in conflict with each other. For college admissions, universities employ either one of the following two policies for students to submit their standardized test scores: applicants freely choosing which scores to submit out of all their scores (super-scoring) and applicants being required to submit all test scores (report-all). The former policy raises equity concerns since lower resourced students may not have the means to take the exam multiple times to increase their chances of receiving a higher test score. From the college's perspective, it is crucial to admit qualified students and reject unqualified students, ensuring the accuracy of the admissions process. Niu et al. conducted an experiment to examine the equity and accuracy scores of the two policies. The results highlighted that the report-all policy has the same effect as requiring students to take the exam only once. From an equity perspective, the report-all policy ensures that students are accepted based on their qualifications regardless of their background. From the college's perspective, the report-all policy helps colleges better identify qualified students compared to that of the super-scoring policy. Therefore, the report-all policy creates an unusual case where accuracy and equity are in alignment with each other~\cite{niu2022best}.

\subsection{Equity versus Social Values }

We identified tensions addressed by researchers between equity and social values such as privacy, inclusivity, and equality. We generally consider social values as norms that are ethical and beneficial to society, organizations or individuals. Examples of social values include justice, equality, freedom, responsibility etc.~\footnote{\url{https://youthfirstinc.org/values-in-todays-society/}}

Holloway illustrated value tensions when creating an inclusive gallery experience for visitors who are visually impaired. Participants included people with vision impairments (PVI), gallery staff members, and artists whose work is being made accessible. While participants raised equity-related values such as inclusion, others addressed values such as artistic integrity (audio descriptions and alternative illustrations are true to the artwork), fiscal responsibility (gallery remains financially viable) and stewardship (ensuring gallery artworks are preserved for future generations) that are paramount to maintaining and preserving the gallery and artworks. The authors discussed the tensions in values such as inclusion and integrity: inclusion requires the artworks translated in different modalities such a touch while integrity assures that the artwork is not misrepresented. While an agreement was not reached by the participants on how to resolve this tension, participants suggested design ideas such as including audio descriptions of the artists having a conversation about the artwork~\cite{holloway2019making}. 

Additionally, Desportes et al. examined how makerspaces could be positioned to address social inequities and support community growth and development. While operationalizing these goals and developing an equity-oriented makerspace, makerspaces could be subject to value tensions. These tensions include favoring individual empowerment versus community impact, openness versus safety, and diversity versus cultural preservation. Regarding the tension between openness of the makerspace and the safety of the participants as well as the equipment in the space, participants suggested ideas such as installing a buzzer system to ensure both openness and safety as well as building a strong sense of community where community members are committed to taking full ownership and responsibility of the space space~\cite{desportes2021examining}.  

So far, the prior examples focus on surfacing stakeholders’ values and value conflicts. However, rather than surfacing value conflicts by impacted stakeholders, Quiñonero-Candela et al. focused on introducing a novel framework that harmonizes equity and equality when designing LinkedIn’s AI system that recommends new job posts to job seekers and potential candidates to recruiters. To achieve harmony of these two values, the framework guides AI practitioners to not only create an AI system that treats everyone equally but also conduct additional efforts to address equity-related issues such as bias, and potential harms. AI practitioners are guided to create an AI system that treats every candidate equally regardless of their demographics and backgrounds. However, the efforts to mitigate biases do not end here. In addition to curating equality, practitioners are guided to achieve an equitable product experience that highlights the users’ experiences, particularly those identified as marginalized~\cite{quinonerocandela2023disentangling}. 

\subsection{Equity versus Equity} 
\label{equity_versus_equity}
Researchers explore how equity can sometimes be in conflict with itself, moving beyond the tensions and trade-offs between equity and other values. One way to operationalize equity could unintentionally create new challenges or inequities. A solution to bring equitable outcomes for one population or stakeholders may unintentionally create inequitable experiences for a different stakeholder group. For example, Solyst et al. discussed the challenges of designing a culturally responsible online STEM education program. Employing a chat functionality between instructors and students can foster equitable experiences for students. A chat system could mitigate the students' discomfort and pressure to turn on one's camera that exposes one's background. This system also encourages quiet students to raise their voices and participate in the program. However, the chat system may not be equitable for everyone, as not every student is proficient in typing. This example shows how a equitable solution may not be equitable in different circumstances~\cite{solyst2022understanding}.  

This tension between a value and itself has already been addressed in fairness literature. Studies highlight fairness definitions can contradict one another, where even using the same dataset may lead to different results~\cite{berk2021fairness,verma2018fairness}. In another instance, algorithms that incorporate protective attributes such as race or gender may exacerbate existing societal inequities, motivating researchers to remove proxies for demographic variables in the dataset~\cite{chohlas-wood2021blind}. However, incorporating protective attributes for disaggregated data, which is data deconstructed to sub-categories, can serve as a method to monitor equity by providing deeper insights on how resources are distributed~\cite{perez2019invisible, madaio2022assessing, mehrabi2021survey}. 
\section{Limitations and Future Work}
Our systematic literature review has two key limitations: a) the choice of our conference venues do not encompass all papers that focus on equity and technology, and b) the choice of keywords and search criteria may exclude relevant papers in our dataset. To address the first limitation, we used ACM-DL “SIGCHI” and “SIGACCESS” as a tool to find HCI papers. Choosing a digital library and relevant venues helped us limit our HCI dataset. We excluded domain-specific venues such as those in education, healthcare, or public policy, where equity discussions are present. Rather, we chose to investigate how equity is discussed broadly within HCI and Fairness. Future work could explore equity within these specific domains to compare how different research communities address equity. As described in Section~\ref{method}, since there is no central database for Fairness-focused venues, we chose relevant conferences to investigate. Our choices were determined by whether the conference focused on the following themes: fairness, accountability, transparency, ethics, equity, and trustworthiness in socio-technical systems and AI. Consequently, we did not investigate venues such as ICML (International Conference on Machine Learning), NeurIPS (Neural Information Processing Systems) that also discuss fairness. Future work could further investigate fairness-related research articles published in these other venues. 
Addressing the second limitation, the search criteria and keyword choice limited our dataset. By constraining our dataset to papers that include "equit*" or "inequit*" in the title, abstract or keyword, we intended to find papers that focus on equity and technology. However, we may have excluded papers that contribute to equity goals in technology but do not explicitly use the word equity. We may also have excluded papers focusing on equity-related themes such as accessibility or social justice, unless they meet our search criteria. For example, Branham et al. surfaced challenges of creating a collaborative and social setting for people who are blind. Despite efforts to surface affected communities’ needs and desires, this paper was not in our dataset since it did not explicitly use the word "equit*" or "inequit*" in the paper~\cite{branham2015collaborative}. Future work could expand upon this limitation and analyze papers that focus on creating equitable outcomes even without "equity" being explicitly mentioned by the authors. 
\section{Discussion}
We observed a significant spike of papers focusing on addressing and integrating equity into technology during the past four years between 2018 and 2022 (see Figure \ref{fig:corpus-year}). Despite this surge of interest, research on equity and technology remains understudied, providing opportunities for future researchers to delve deeper into its complexities and nuances.
Based on our literature review, we propose a synthesis of the findings through an equity framework and discuss how future researchers who aim to study the intersection of equity and technology can borrow and extend different dimensions of this framework.
These dimensions include: 1) \textbf{What} shapes equity research, 2) \textbf{Why} study equity and technology, 3) \textbf{How} do researchers conduct equity research, and 4) 
\textbf{Tensions \& Trade-offs} in advancing equity. Additionally, we encourage researchers to reflect on technology as a solution. In this section, we provide 7 recommendations and key factors that researchers can consider for each of the dimensions. 

\begin{figure}[p]
    \centering
    \includegraphics[width=0.95\linewidth]{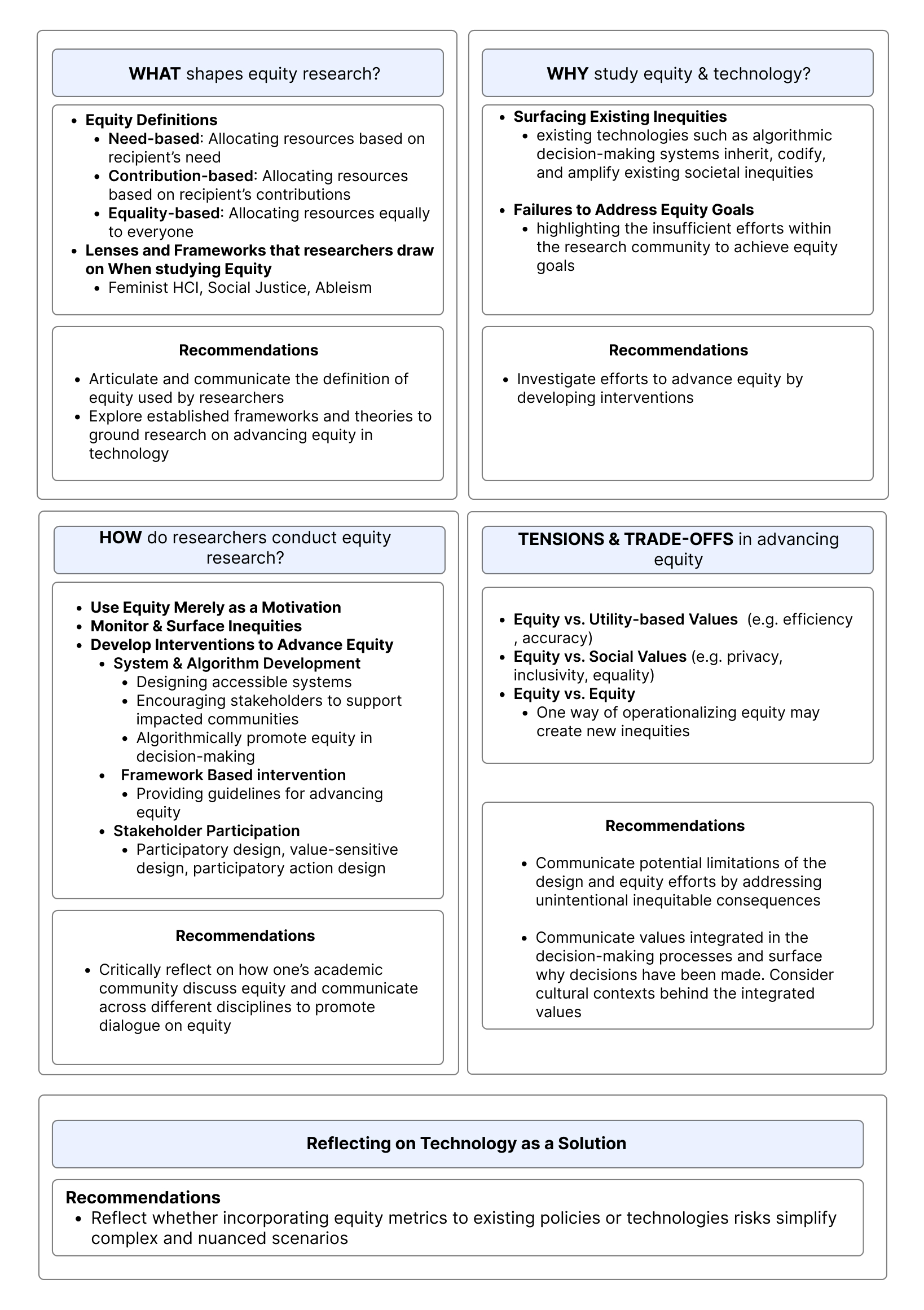}
    \caption{A Framework for future researchers aiming to study the intersection of equity and technology. This framework illustrates the key dimensions for researchers conducting research in equity and technology. In addition to the key dimensions, we elaborate guidelines and recommendations for researchers to reflect and consider.} 
    \label{fig:equity_framework}
\end{figure}

\subsection{WHAT: Definitions and Lenses to Shape Equity Research}
\subsubsection{Providing a Clear and Explicit Equity Definition}
We discovered a significant gap in the clarity and consistency of equity definitions among researchers. In our review, we identified three primary definitions used by researchers (need-based definition, contribution-based definition, and equality-based definition). 
Although these three definitions may not encompass all existing concepts of equity, these serve as a starting point when articulating equity. We do not believe that HCI and Fairness require a single definition of equity, and researchers should not limit their discussions and definition of equity to only these three definitions that we defined.

While we identified these three primary definitions used by researchers, it was striking that a substantial proportion of papers did not explicitly define equity at all. This absence of a clear definition of equity is more pronounced in HCI research compared to papers in the Fairness-focused venues (Only 21\% of HCI papers provide a definition, while this was 40\% for the Fairness-focused venues). This disparity in defining equity between the fairness and HCI communities could be attributed to the greater emphasis on algorithm and system development within the fairness community, where equity is often expressed through computations. These efforts to promote equity typically necessitate a clear definition of the concept to effectively evaluate progress and measure outcomes. 

Given these diverse definitions of equity, we suggest scholars to articulate their choice of equity definition, using these three definitions as a starting point.
However, the lack of explicit and clear equity definition could create challenges for effective communication and collaborations among researchers. These efforts to clearly state the researchers' own definition of equity will enable researchers to build upon each other's work effectively and contribute to a coherent body of knowledge.
\\
\\
\textit{\textbf{(Recommendation 1) Articulate and communicate the definition of equity used by the researchers. Use need-based, contribution-based, and equality-based definitions of equity as a starting point but do not limit to only these three definitions}} 

\subsubsection{Identifying the Related Lenses and Frameworks When Studying Equity}
When investigating equity, we suggest researchers to consider employing the lenses and frameworks that are relevant and connected with equity. In our corpus, we found researchers employing Feminist HCI, social justice, and ableism as means to discuss equity (Section~\ref{section5_lenses_framework}). However, these are only examples of lenses and frameworks that could be applied when conducting equity research. Fairness is another theme that could be employed for equity research, surfacing impacted stakeholders' perceptions of fairness, as well as measuring and evaluating fairness (Section~\ref{relatedwork}). Other lenses and frameworks relevant to equity but not have been explored in this literature review are critical race theory~\cite{ogbonnaya2020critical}, intersectionality~\cite{wisniewski2018intersectionality}, and queer theory~\cite{taylor2024cruising}. 
\\
\\
\textit{\textbf{(Recommendation 2) Investigate and explore established frameworks and theories that could ground research on advancing equity in technology}} 

\subsection{WHY: Going Beyond Bias Mitigation} 
In Section~\ref{section6:why}, we surface how researchers discuss equity and technology to either surface existing inequities or address the research community's lack of attention to equity goals. Researchers were often motivated by two different types of inequities: societal inequities and inequitable behaviors of technologies. Due to societal discrimination and structural inequities, researchers have been motivated to understand barriers of marginalized communities and how technologies may meet their needs. However, the existence of technology itself may further exacerbate existing societal inequities, motivating researchers to surface how these technologies affect impacted stakeholders. In addition to these inequities in society and technology, researchers articulate the lack of focus to advance equity within the research community, inspiring them to investigate inequities beyond mitigating existing biases (Section~\ref{section6:6.2_failures}).

Despite these motivations from researchers who invest in equity research, most papers in our corpus merely discussed equity as a motivation to their work and only aimed to monitor and surface inequities (Table~\ref{tab:focus}). This trend surfaced in both the HCI and Fairness communities. While monitoring and surfacing inequities are important as a first step to addressing inequities, we recommend researchers go beyond taking an reactive approach to mitigating biases. Researchers could take a proactive approach to equity challenges by not only reflecting on how they are using equity to motivate their work but also evaluating whether their solutions and findings contribute to advancing equity that aligns with their motivations.  
\\
\\
\textit{\textbf{(Recommendation 3) Investigate efforts to advance equity by developing interventions using Figure~\ref{fig:equity_framework} and Table~\ref{tab:focus} as a starting point for researchers to find the suitable method for their research. }}

\subsection{HOW: Bridging Different Efforts in Fostering Equity}

Our study highlights differences in how HCI and the Fairness community approach in promoting equity in technology (Table~\ref{tab:focus}). We illustrated the extent to which equity is discussed as the focus of the research and design, as well as how researchers develop efforts to foster or integrate equity in technology. Researchers may discuss equity merely as a motivation, monitor and surface biases and inequities, or develop interventions to advance equity. 

We found that the HCI community tends to surface existing inequities through qualitative methods such as interviewing stakeholders, evident in 47\% of the papers. On the other hand, the Fairness community focuses on building systems/algorithms (39\%) and frameworks (38\%). While there is no correct approach to foster equity, limitations exist for each method. In the case of the Fairness-focused venues, scholars often develop equity heuristics without involving impacted stakeholders. Due to the complex nature of equity, the heuristic developed by researchers may not be considered equitable for impacted communities and may create unintended harms. On the other hand, solely surfacing existing inequities experienced by impacted stakeholders may not lead to any concrete interventions or systems that create equitable outcomes. Without any concrete equity assessments of the design implications suggested by researchers, it is challenging to discuss the designs’ impact on equity. 

We acknowledge that scholars who reside in the intersection between the HCI and Fairness community may be conducting different research methods or approaches when addressing equity depending on the venue, audiences of the research community, as well as the venues' methodological preferences. According to our analysis in Section~\ref{section4:authors-keywords-context}, we found only 6 authors (538 unique authors in the HCI corpus, 311 unique authors in the Fairness corpus) who published at both HCI and Fairness-focused venues in our corpus. While our analysis did not consider the degree of influence the researchers had on the choice of the methodology of their papers, we recognized a small overlap of researchers in the intersection between the HCI and Fairness-focused venues when discussing equity. We also found that it is often the case where these scholars employed similar research methods regardless of the venue.

Nevertheless, the differences in disciplines and methodologies across the two communities may be intentional, given the diverse audiences for each venue. Different research approaches tackling equity challenges contribute to diverse and novel solutions. While such diverse perspectives on addressing inequities are valuable, communication of these differences is paramount. As discussed in our findings, equity is a complex and nuanced concept with a lack of concrete definition; therefore, scholars should build upon each other’s work and motivate each other to hopefully merge the strengths of different disciplines.   
While acknowledging these differences in discipline and focus across research communities and venues, we believe that having awareness of how different venues address equity concerns may help researchers to critically reflect on their own practices and assumptions about equity. Therefore, we suggest the two communities have more communications on best practices to integrate equity to technologies. By developing academic workshops and panels focused on equity and technologies, we encourage scholars from both communities to share their findings, methods, and experiences to initiate conversations and further advance research on integrating equity in technologies~\cite{yoo2023beyond}.  

In addition to bridging the gap between the HCI and Fairness communities, we encourage researchers to critically reflect on their academic community and conferences, surfacing how equity is being discussed and communicated. In our corpus, we found only four SIGACCESS papers that fit our inclusion and exclusion criteria. It is possible that the SIGACCESS community uses different terminologies related to equity, such as accessibility, inclusivity, and social justice, or employs diverse methods such as universal design~\cite{persson2015universal}. Equity discussions are indeed present within the SIGACCESS community; however, they may not be framed in the same manner as in other research communities. This variety in how equity is addressed highlights the need for research communities to explicitly clarify their definitions and approaches when integrating and advancing equity. These practices could further promote dialogue on the topic from various perspectives.
\\
\\
\textit{\textbf{(Recommendation 4) Critically reflect on how one's academic community and conferences discuss equity. Communicate across different disciplines to promote dialogue on equity. }} 

\subsection{ TENSIONS \& TRADE-OFFS: Enhancing  Value Transparency in Advancing Equity}
In Section~\ref{section8:tensions}, our results highlight the complexities and nuances of integrating equity within technology, particularly when it comes to navigating the tensions and trade-offs between equity and other values such as utility-based values, social values and even equity itself (Figure~\ref{fig:equity_framework}). A critical aspect of this challenge is the realization that equitable solutions favoring one stakeholder group may inadvertently result in inequities for others, leading to unintended negative consequences~\cite{solyst2022understanding}.

These findings align with Value Sensitive Design (VSD), a theoretically grounded methodology aimed to incorporate stakeholders’ values into the technology's design process~\cite{friedman1996value}. Previously, scholars investigated value tensions arising from misinterpretations of values~\cite{manders2011values}, and the dissonance between the values of impacted stakeholders and the values operationalized in practice~\cite{robertson2021modeling}. Therefore, it is crucial for researchers to critically reflect on the potential risks and unintended impacts of their design solutions and interventions for equity integration into technology. However, our review revealed a scarcity of such critical reflection in the researcher papers. 
\\
\\
\textit{\textbf{(Recommendation 5) Communicate potential limitations of the design and equity efforts by addressing unintentional inequitable consequences as well as tensions and trade-offs between equity and other values. }}
\\
\\
Surfacing and addressing unintentional inequitable outcomes involve a more nuanced approach to transparency in equity research, particularly in terms of value conflicts and decisions. We suggest researchers aiming to navigate this problem to consider the different tensions and trade-offs described in Figure~\ref{fig:equity_framework} as an initial reflection on how their approach may conflict with other values. Moreover, building on the concept of “value transparency” introduced by Park et al.~\cite{park2022power}, we see an opportunity for researchers to not only communicate how decisions are made but also more crucially, reflect \textit{why} certain decisions are made over others. Value transparency involves articulating the rationale behind prioritizing or comprising certain values, which is especially pertinent in the context of equity where multiple conflicting values may be at play. This approach promotes open discussions, enabling the research community to navigate the trade-offs and complexities inherent in balancing equity with other competing values. 

Moreover, in Section~\ref{equity_versus_equity}, we found equity in conflict with equity, particularly as one way of integrating equity into technology may create inequitable outcomes. Therefore, when addressing values, we suggest researchers add the cultural contexts and perspectives in which the values were created rather than making implicit claims into the design~\cite{borning2012next}. Consequently, such transparency can lead to more informed, thoughtful, and inclusive technology designs that more effectively address the multifaceted nature of equity. 
\\
\\
\textit{\textbf{(Recommendation 6) Communicate values integrated in the decision-making processes and surface why certain decisions have been made. Integrate the cultural contexts behind the values to address the nuanced nature of equity.}} 

\subsection{Reflecting on Technology As a Solution: The Perils of Techno-Solutionism in Developing Equitable Solutions} 
Is developing a technology the best solution for creating equitable outcomes? In this section, we encourage researchers to reflect on whether their design recommendations and solutions intended to create equitable outcomes inherently address existing inequities. With technologies often exacerbating existing societal inequities and biases, there are efforts to mitigate bias and create equitable solutions by developing new technologies or making modifications to existing ones - for example, auditing algorithms~\cite{devos2022toward,shen2021everyday} or deploying new technologies that identify risks and biases~\cite{veale2018fairness}. This tendency to resolve challenging problems with technology is known as techno-solutionism~\cite{morozov2013save}. 

However, researchers found that creating new technologies may not always be the solution that impacted stakeholders want and need. For instance, in the case of child-welfare predictive algorithms, Stapleton et al. surfaced that impacted stakeholders, particularly parents, suggested low-tech or no-tech alternatives to predictive systems. They instead advocated for improvements in the hiring, training and the child welfare team’s decision-making process without the use of any technologies~\cite{stapleton2022imagining}. In the context of public sector organizations, adopting algorithms may lead to over-simplifying complex and nuanced government policies and reducing them to mathematical metric to fit into the algorithm design~\cite{levy2021algorithms}. 

As researchers develop equity interventions or suggest new design implications to promote equity, we encourage scholars to re-evaluate whether developing a technology inherently resolves equity issues. 
\\
\\
\textit{\textbf{(Recommendation 7) Reflect whether incorporating equity metrics to existing policies or technologies risks simplify complex and nuanced scenarios. }}

\section{Conclusion}
Through our systematic literature review, we provide an overview of 202 research papers addressing equity and its integration into technology across HCI and selected Fairness-focused venues from their inception to 2023. We found a rapidly growing interest in addressing inequities and developing interventions to promote equity in technologies during the past four years. Researchers surface tensions and trade-offs between equity and other values, highlighting the complexities of integrating equity in technologies. Moreover, we see a pattern of researchers using existing equity-related frameworks and lenses to reflect on designs that address inequities. However, there still remain important gaps that need to be addressed. Inconsistencies in how equity is defined and the lack of explicit equity definitions call for the need to clearly define equity. Equity is a nuanced and context-dependent concept which requires a critical reflection on whether design or technological solutions may lead to equitable outcomes. Our review reveals a considerate number of papers briefly addressing equity without diving deeper into the concept. As there is still a dearth of research focusing on equity, we encourage researchers to actively invest in this topic and share findings and methods across different disciplines.

\begin{acks} 
The research was funded by the Carnegie Mellon University Block Center for Technology and Society Award No. 55303.1.5007719 and the National Science Foundation (NSF) under Award No. 1952085. We appreciate the anonymous reviewers for their invaluable feedback on improving the paper. 
\end{acks}

\bibliographystyle{ACM-Reference-Format}
\bibliography{reference}

\appendix
\appendix

\clearpage
\section{Authors} 
\label{appendix:authors}
\begin{table*}[h!]
  \centering
  \caption{The table shows the authors who published to both HCI and Fairness-focused venues within our corpus.}
    \begin{tabular}{p{0.2\textwidth} p{0.8\textwidth}}
        \toprule
        \textbf{Authors} & \textbf{Papers} \\
        \midrule
        Bennett, Cynthia L. & 
        “It’s Complicated”: Negotiating Accessibility and (Mis)Representation in Image Descriptions of Race, Gender, and Disability; \newline
        AI’s Regimes of Representation: A Community-centered Study of Text-to-Image Models in South Asia \\
        \midrule
        Bidwell, Nicola J. & 
        Invigorating Ubuntu Ethics in AI for healthcare: Enabling equitable care; \newline
        Peer-to-peer in the Workplace: A View from the Road; \newline
        From Treatment to Healing: Envisioning a Decolonial Digital Mental Health \\
        \midrule
        Bruckman, Amy & 
        Toward a Grassroots Culture of Technology Practice; \newline
        Mitigating Racial Biases in Toxic Language Detection with an Equity-Based Ensemble Framework \\
        \midrule
        Chouldechova, Alexandra & 
        Algorithmic Fairness and Vertical Equity: Income Fairness with IRS Tax Audit Models; \newline
        Fairness in Risk Assessment Instruments: Post-Processing to Achieve Counterfactual Equalized Odds; \newline
        The Impact of Algorithmic Risk Assessments on Human Predictions and its Analysis via Crowdsourcing Studies \\
        \midrule
        Harrington, Christina & 
        Deconstructing Community-Based Collaborative Design: Towards More Equitable Participatory Design Engagements; \newline
        Envisioning Equitable Speech Technologies for Black Older Adults; \newline
        Examining Identity as a Variable of Health Technology Research for Older Adults: A Systematic Review; \newline
        Speculative Blackness: Considering Afrofuturism in the Creation of Inclusive Speculative Design Probes \\
        \midrule
        Kingsley, Sara & 
        "Give Everybody [..] a Little Bit More Equity": Content Creator Perspectives and Responses to the Algorithmic Demonetization of Content Associated with Disadvantaged Groups; \newline
        Narratives and Counternarratives on Data Sharing in Africa \\
        \bottomrule
    \end{tabular}
\end{table*}

\clearpage
\section{Author Keywords}
\label{appendix:author-keywords}
\begin{table*}[h]
  \centering
  \caption{The table shows the top author keywords for HCI and Fairness venues. We present the 10 keywords that were mentioned more than twice. We list the top 14 keywords in HCI as multiple keywords that were addressed 5 times. In Fairness community, we list 6 keywords that were addressed more than twice.}
  \label{tab:keywords}
   \begin{tabular}{p{0.2\textwidth} p{0.75\textwidth}}
        \toprule
        \textbf{Venues} & \textbf{Top Keywords and Count} \\
        \midrule
        \textbf{HCI} (Top 14) & equity (12), participatory design (12), gender (12), accessibility (10), feminist hci (6), collaboration (5), children (5), ictd (5), design (5), race (5), social media (5), crowdsourcing (5), hci (5), older adults (5) \\
        \textbf{Fairness} (Top 6) & fairness (9), algorithmic fairness (8), machine learning (4), equity (3), ethics (3), accountability (3) \\
        \bottomrule
    \end{tabular}
\end{table*}

\section{Multi-Methods Distribution}
\label{appendix:methods-details}
\begin{table*}[h]
  \centering
  \caption{Distribution of mixed method used in HCI venues.}

  \label{appendix:multimethod-hci}
  \begin{tabular}{p{0.5\textwidth} p{0.3\textwidth}}
        \toprule
        \textbf{Methods} & \textbf{Number of Papers in HCI Using Multiple Methods} \\
        \midrule
        Qualitative; Quantitative & 15 \\ 
        Design; Qualitative & 15 \\ 
        Technical/Systems; Quantitative & 8 \\ 
        Technical/Systems; Qualitative; Quantitative & 7 \\ 
        Design; Technical/Systems; Qualitative & 7 \\ 
        Technical/Systems; Qualitative & 6 \\ 
        Design; Qualitative; Quantitative & 3 \\ 
        Theoretical; Qualitative; Design & 2 \\ 
        Theoretical; Qualitative & 2 \\ 
        Design; Theoretical; Qualitative & 1 \\ 
        Theoretical; Qualitative; Quantitative & 1 \\ 
        \midrule
        \textbf{Total} & \textbf{67 / 141} \\
        \bottomrule
    \end{tabular}
\end{table*}

\begin{table*}[h]
  \centering
  \label{appendix:multimethod-fairness}
\caption{Distribution of mixed method used in Fairness venues.}
 \begin{tabular}{p{0.5\textwidth} p{0.3\textwidth}}
    \toprule
    \textbf{Methods} & \textbf{Number of Papers in Fairness Using Multiple Methods} \\
    \midrule
    Technical/Systems; Quantitative & 15 \\ 
    Technical/Systems; Theoretical; Quantitative & 11 \\ 
    Theoretical; Quantitative & 6 \\ 
    Theoretical; Qualitative & 3 \\ 
    Technical/Systems; Design; Qualitative & 2 \\ 
    Design; Qualitative & 1 \\ 
    Theoretical; Qualitative; Quantitative & 1 \\ 
    Design; Qualitative; Quantitative & 1 \\ 
    Technical/Systems; Theoretical & 1 \\ 
    Qualitative; Quantitative & 1 \\ 
    \midrule
    \textbf{Total} & \textbf{42 / 61} \\
    \bottomrule
\end{tabular}
 
\end{table*}

\end{document}